\documentclass[aps,showpacs,prapplied,twocolumn,superscriptaddress,floatfix]{revtex4-2}
\usepackage[colorlinks,urlcolor=blue,linkcolor=blue,anchorcolor=blue,citecolor=blue]{hyperref}
\makeatletter
\usepackage[utf8]{inputenc}
\usepackage{graphicx,bm,color}
\usepackage[dvipsnames]{xcolor}
\usepackage[margin=1in]{geometry}
\usepackage{amsthm,amsmath,amssymb,mathtools}
\usepackage{multirow}
\usepackage{titlesec}
\titlespacing*{\section}
{0pt}{2\baselineskip}{\baselineskip}
\titlespacing*{\subsection}
{0pt}{\baselineskip}{\baselineskip}
\usepackage{float}
\let\newfloat\newfloat@ltx
\makeatother
\usepackage{algorithm,algpseudocode}
\usepackage[labelfont=bf,justification=Justified,font=small]{caption}
\usepackage{cleveref}
\usepackage[caption=false]{subfig}
\usepackage{stackengine,scalerel}
\usepackage{soul}

\newcommand{\rbracket}[1]{\left(#1\right)} %round bracket
\newcommand{\sbracket}[1]{\left[#1\right]} %square bracket
\newcommand{\cbracket}[1]{\left\{#1\right\}} %curly bracket
\newcommand{\noq}{n}
\newcommand{\noH}{10}

\newcommand{\costfunction}{\text{C}}

 \theoremstyle{plain}
 
 \theoremstyle{plain}
 
 \theoremstyle{plain}
 
 \theoremstyle{plain}
  
 \theoremstyle{plain}
 
 \theoremstyle{plain}
 
 \theoremstyle{plain}
 
 \theoremstyle{remark}
 \newtheorem*{rem*}{Remark}
 \theoremstyle{plain}
  \newtheorem{rem}{Remark}

\theoremstyle{plain}
 \newtheorem*{conj*}{Conjecture}
 \theoremstyle{plain}

\usepackage{comment}

\usepackage[normalem]{ulem} %for \sout

\input{Qcircuit}

\definecolor{teal}{HTML}{00B5AD}

\begin{document}

\title{Design and execution of quantum circuits using tens of superconducting qubits and thousands of gates for dense Ising optimization problems}
\author{Filip B. Maciejewski}
\author{Stuart Hadfield}
\affiliation{Research Institute for Advanced Computer Science (RIACS), USRA, Moffett Field, CA}
\affiliation{Quantum Artificial Intelligence Laboratory (QuAIL), NASA, Moffett Field, CA}
\author{Benjamin Hall}
\affiliation{Research Institute for Advanced Computer Science (RIACS), USRA, Moffett Field, CA}
\affiliation{Quantum Artificial Intelligence Laboratory (QuAIL), NASA, Moffett Field, CA}
\affiliation{Michigan State University, East Lansing, MI}
\author{Mark Hodson}
\author{Maxime Dupont}
\author{Bram Evert}
\affiliation{Rigetti Computing, Berkeley, CA}
\author{James Sud}
\author{M. Sohaib Alam}
\author{Zhihui Wang}
\affiliation{Research Institute for Advanced Computer Science (RIACS), USRA, Moffett Field, CA}
\affiliation{Quantum Artificial Intelligence Laboratory (QuAIL), NASA, Moffett Field, CA}
\author{Stephen Jeffrey}
\author{Bhuvanesh Sundar}
\affiliation{Rigetti Computing, Berkeley, CA}
\author{P. Aaron Lott}
\affiliation{Research Institute for Advanced Computer Science (RIACS), USRA, Moffett Field, CA}
\affiliation{Quantum Artificial Intelligence Laboratory (QuAIL), NASA, Moffett Field, CA}
\author{Shon Grabbe}
\affiliation{Quantum Artificial Intelligence Laboratory (QuAIL), NASA, Moffett Field, CA}
\author{Eleanor G. Rieffel}
\affiliation{Quantum Artificial Intelligence Laboratory (QuAIL), NASA, Moffett Field, CA}
\author{Matthew J. Reagor}
\affiliation{Rigetti Computing, Berkeley, CA}
\author{Davide Venturelli}\email{dventurelli@usra.edu}
\affiliation{Research Institute for Advanced Computer Science (RIACS), USRA, Moffett Field, CA}
\affiliation{Quantum Artificial Intelligence Laboratory (QuAIL), NASA, Moffett Field, CA}

\begin{abstract}
We develop a hardware-efficient ansatz
for variational optimization, derived from existing ans\"atze in the literature, that parametrizes subsets of all interactions in the Cost Hamiltonian in each layer. 
We treat gate orderings as a variational parameter and observe that doing so can provide significant performance boosts in experiments. 
We carried out experimental runs of a compilation-optimized implementation of fully-connected Sherrington-Kirkpatrick Hamiltonians on a 50-qubit linear-chain subsystem of Rigetti's Aspen-M-3 transmon processor. 
Our results indicate that, for the best circuit designs tested, the average performance at optimized angles and gate orderings increases with circuit depth (using more parameters), despite the presence of a high level of noise. 
We report performance significantly better than using a random guess oracle for circuits involving up to $\simeq5,000$ two-qubit and $\simeq5,000$ one-qubit native gates. We additionally discuss various takeaways of our results toward more effective utilization of current and future quantum processors for optimization.
\end{abstract}
\maketitle

\begin{figure*}[!t]
\includegraphics[width=1.0\textwidth]{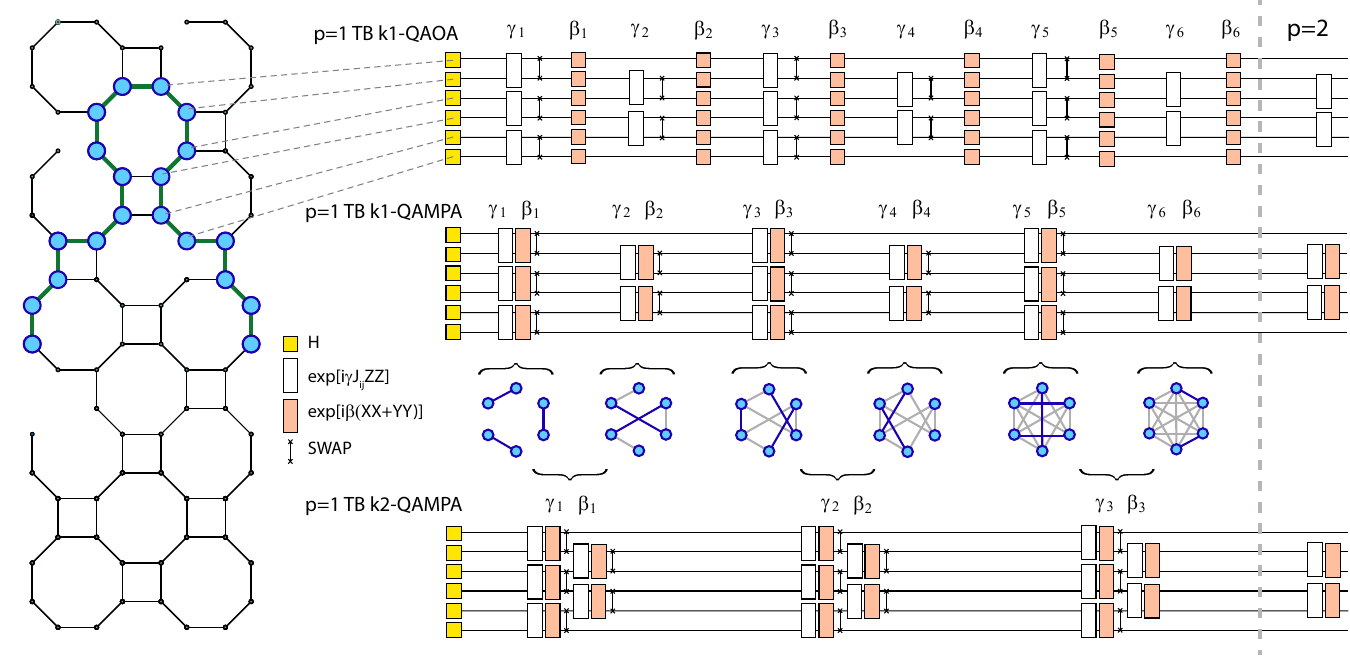}     \caption{
Left: a subset of $n=20$ qubits with linear connectivity on the physical Aspen-M-3 chip device. The presented subset is an example; the actual linear chains used in our experiments were chosen based on calibration data, see Appendix~\ref{app:expDetails}. Right: circuits implementing 6-qubit \textit{Time-Block} (TB) k1-QAOA, k1-QAMPA and k2-QAMPA ans\"atze. Circuit gates are Hadamards (yellow), Phase-Separator $ZZ_{i,j}^{(J)}$ gates (white) or mixing gates (red, either $X$ rotations or $XY$ operators) and SWAPs. 
The pictorialized graphs illustrate the progressing scheduling of the interactions as the circuit and the swap network is executed up to the realization of a complete graph (which marks the execution of a \emph{full algorithmic round}). After $p=1$, the round is repeated but mirrored for compilation efficiency (only the first gates of the next round are drawn). } \label{fig:paper_overview}

\end{figure*}

\section{Introduction}\label{sec:introduction}

As Noisy Intermediate-Scale Quantum (NISQ) processing hardware relentlessly improves and crosses the tractability threshold of exact classical simulation~\cite{arute2019quantum, Zhong2020quantum,Wu2021strong,Madsen2022quantum,morvan2023phase,Kim2023evidence}, many questions remain with regard to translating such behavior to improved performance for real-world problems.
In the NISQ context, the task of quantum optimization consists of designing and validating software protocols executable on pre-fault-tolerant quantum processors to output high-quality solutions to instances of a challenging combinatorial problem.
While a series of initial experimental demonstrations are encouraging~\cite{harrigan2021quantum,otterbach2017unsupervised,qiang2018large,abrams2019implementation,bengtsson2019quantum,willsch2020benchmarking,pagano2020quantum,niroula2022constrained,ebadi2022quantum,nguyen2023quantum,shaydulin2023qaoa,dupont2023quantum,pelofske2023high,pelofske2023scaling}, there is still no clear pathway to tackle the detrimental role of noise and other hardware limitations (connectivity, native gate sets, etc.), in a regime where the use of a quantum solver would be preferable to classical alternatives.
In this paper, we ask a related question: given a real-world device with limited qubits, connectivity and gate primitives, how can it be effectively utilized to solve optimization problems? 
The severe limitation in the number of logical layers that can be reliably used in algorithms with good fidelity has motivated the community to develop and test hardware-efficient ans\"atze, 
~\cite{kandala2017hardware,tang2021qubit,leone2022practical,d2023challenges}, i.e., variational algorithms that require minimal resources overhead at the expense of increased parametrization. 
That is often traded off with an inferior understanding of the ability of these approaches to find good solutions (even in principle), and generally more challenging task of parameter setting.
A related approach would be to take existing algorithms that have some theoretical guarantees such as the Quantum Alternating Operator Ansatz (QAOA) ~\cite{farhi2014quantum,hadfield2019quantum,dalzell2023quantum,abbas2023quantum} or the Quantum Alternate Mixer-Phaser Ansatz (QAMPA)~\cite{larose2022mixer}, and derive hardware-adapted variants that exploit the problem structure in order to use as little resources as possible of the quantum hardware. 
We take this approach by presenting in this paper three main contributions:
\begin{itemize}
    \item Designing novel ans\"atze optimized to execute on quantum gate-model processors that feature a linear chain of connected qubits as a sub-graph of its chip layout, targeting random fully-connected graph optimizations.
    \item Treating gate ordering of the ansatz as a variational parameter, thus allowing for a more fine-grained control over the circuit structure. This allows for significant performance boosts in experiments.
    \item Implementing and executing our circuit variants on linear chains of qubits of Aspen-M-3 quantum processing unit (QPU) by Rigetti Computing, up to sizes of $n=50$ qubits and using up to $\simeq 5,000$ two-qubit and $\simeq 5,000$ one-qubit gates, which, to the best of our knowledge, is a complexity regime beyond what has been reported so far in the applied quantum optimization literature with gate-model devices. 
\end{itemize} 

The main experimental goal of our paper is twofold. 
First, we compare the optimization performance of our Time-Block ansatze with standard variants of QAMPA and QAOA.
Results indicate that, for our best-performing circuit design (the Time-Block QAMPA ansatz), the average performance at optimized parameters increases with algorithm depth (with each depth unit adding $O(\noq)$ gates to the circuit), and it performs overall better than vanilla QAMPA. For QAOA we do not observe a significant advantage of the Time-Block strategy compared to the vanilla variant.
Second, we compare the experimental results to a random guess solver (equivalent to a strong white noise regime) given the same number of function calls, and we observe performance gains over this simple classical strategy.

An illustrated overview of the ideas presented in this paper can be found in Fig.~\ref{fig:paper_overview}. Our main experimental results are presented in Fig.~\ref{fig:best_results}, showing a pictorial view of the implementation of our ansatz on fully-connected random Sherrington-Kirpatrick Hamiltonians.

The paper is structured as follows: in the next section (\ref{sec:preliminaries}) we recap prior work, define the optimization problem and the two key ans\"atze (QAOA and QAMPA) as well as their efficient compilation using swap networks and native gates calibrated on Aspen-M-3. 
In section \ref{sec:ansatze} we introduce new types of QAOA and QAMPA circuits that we call \emph{Time-Block} (TB) k-QAOA/QAMPA (with the parameter $k$ controlling depth of a single layer).
Section \ref{sec:methodology} will discuss the experimental methodology for the benchmarks and in Section~\ref{sec:results} we present the results of the optimization performance for different system sizes, algorithm depth, and parameter choices. 
In Section \ref{sec:discussion} and \ref{sec:conclusion} we discuss the results conclude with considerations for future improvements.
More details on the experiments and data are made available in the Appendix.

\section{Preliminaries}\label{sec:preliminaries}

Quantum ans\"atze design is known to exhibit various critical tradeoffs between performance, the number of parameters, and the difficulty in optimizing them, with numerous strategies and heuristics proposed \cite{blekos2023review}. 
At a high level, parameter setting can become hard due to the curse of dimensionality, as well as difficulties arising from the cost function landscapes such as barren plateaus~\cite{mcclean2018barren,cerezo2021cost,wang2021noise,ragone2023unified} or 
plentiful local minima~\cite{wierichs2020avoiding,you2021exponentially,anschuetz2022quantum}. 
At the low level, a distinct but related concern is how to best utilize a fixed quantum hardware device, which will naturally be limited in terms of the number of qubits, their connectivity, coherence time, and gate fidelities, among other important properties. 
The latter perspective especially has motivated the development of hardware-efficient ans\"atze (i.e., most operations do not need to be compiled~\cite{kandala2017hardware,farhi2017quantum,tang2021qubit,leone2022practical,d2023challenges}) with reduced quantum resource requirements. 
A natural approach to constructing hardware-efficient ans\"atze is to build them from the native gate set. 
To increase expressibility, one may take a low-depth realization or variant of an existing algorithm such as QAOA, and increase the number of free parameters.
However, this
generally comes at the cost of 
a more challenging 
parameter optimization
~\cite{mcclean2018barren,cerezo2021cost,nakaji2021expressibility,uvarov2021barren,bittel2021training,holmes2022connecting,napp2022quantifying,zhang2022fundamental,larocca2023theory}.
In the next section, we present a hardware-efficient ansatz design developed using this methodology.

\subsection{QAOA and QAMPA}\label{sec:ansatze_circuits}
We start by briefly reviewing the main details of QAOA \cite{farhi2014quantum,hadfield2019quantum,dalzell2023quantum,abbas2023quantum} and QAMPA \cite{larose2022mixer}. 
 In Sec.~\ref{sec:ansatze} we will 
 build off of these approaches to derive new hardware-efficient ans\"atze. 
In what follows, $X_i,\ Y_i,\ Z_i$ denote the respective Pauli matrices acting on $i$th qubit.

In each ansatz,
a sequence of $p$ parameterized circuit layers is applied to an initial state $\ket{\psi_0}$ for $n$ qubits. 
While different choices are possible, we consider  $\ket{\psi_0} = \ket{+}^{\otimes \noq}$ as the initial state for both ans\"atze ($\ket{+}$ denoting the $+1$ eigenstate of the $X$ operator).
For standard QAOA 
each layer has the structure 
\begin{align}\label{eq:qaoa_layer}
(\text{QAOA layer}) \quad U_B\rbracket{\beta}U_{C}\rbracket{\gamma} \ ,
\end{align}
where $U_B\rbracket{\beta} = \bigotimes_{i=1}^{n}\exp\rbracket{-i \beta \ X_i}$ is the mixing operator and $U_{C}\rbracket{\gamma} = \exp\rbracket{-i \gamma\ H_C}$ is the phase separator, with $H_C$ the problem Hamiltonian, specified by real coefficients $(J)\equiv\{J_{i,j}\}$ (see Eq.~\eqref{eq:fully_connected_hamiltonian}). $\gamma$ and $\beta$ are real angle parameters. 
For implementation, QAOA operators are typically compiled to quantum circuits consisting of one- and two-qubit gates. 
 
For QAMPA, the ansatz looks significantly different -- the mixer and phase separator are altered to become jointly locally implemented for each pair of qubits. As discussed in \cite{larose2022mixer}, this ansatz can be considered an approximation of QAOA with XY mixers.
A single layer of QAMPA has the following structure
\begin{align}\label{eq:qampa_layer}
    (\text{QAMPA layer}) \quad \prod_{\rbracket{i,j}\in \mathcal{S}}  \ 
    \tilde{u}_{i,j}\rbracket{\gamma,\beta}
\end{align} where 
\begin{equation}
\begin{split}
    &\tilde{u}_{i,j}\rbracket{\gamma,\beta} \coloneqq     
    ZZ_{i,j}^{(J)}\rbracket{\gamma}\  XY_{i,j}\rbracket{\beta} \ , \\
    &ZZ_{i,j}^{(J)}\rbracket{\gamma} = \exp\rbracket{-i\gamma J_{i,j} Z_iZ_j} \ ,\\
        &XY_{i,j}\rbracket{\beta} = \exp\rbracket{-i\beta \rbracket{X_iX_j+Y_iY_j}} \ ,\\
\end{split}
\end{equation}
and $\mathcal{S}$ is some \textit{ordered} set of pairs of indices.
Note that since the $XY$ operators on overlapping pairs of qubits do not commute, different gate orderings $\mathcal{S}$ implement different ans\"atze in general.

We remark that the QAMPA ansatz has a symmetry that conserves Hamming weight of bitstrings, and thus does not mix subspaces of computational basis with different Hamming weights of the support.
For the class of Sherrington-Kirkpatrick model instances we consider below (see Eq.~\eqref{eq:fully_connected_hamiltonian}) drawn with equal probabilities of positive or negative edge weights, we intuitively expect near-optimal solutions to concentrate close to Hamming weight $\noq/2$ ($\noq$ being a number of variables), which is the region where most of the support of $\ket{\psi_0}$ lies.
For applications that observe similar Hamming weight concentration, adoption of the QAMPA ansatz may be desirable~\footnote{It should be noted that a real-world NISQ version of the algorithms does not conserve the symmetry due to noise and miscalibrations - which might be occasionally an advantage versus noiseless execution.}.

\subsection{Problem overview, results, and algorithms}

Our target problem is the Sherrington-Kirkpatrick (SK) model \cite{sherrington1975solvable}, a combinatorial minimization problem on complete graphs with random couplings.  
QAOA for the SK model was previously considered in \cite{farhi2022quantum, basso2021quantum, boulebnane2021predicting,marwaha2022bounds}. 
Problem instances are 
specified by the fully-connected Ising Hamiltonian on $\noq$ quantum spins
\begin{eqnarray}\label{eq:fully_connected_hamiltonian}
H_C &=& \sum_{i<j}^{\noq} J_{i,j} Z_iZ_j,\label{eq:H_C-SK}
\end{eqnarray}
where $J_{i,j}$ are drawn uniformly at random from $\{-1,+1\}$.

The SK model has been the subject of a long history of scientific investigations, and while it could be considered a solved model from the point of view of thermodynamics~\cite{panchenko2012sherrington} (i.e., as $\noq\rightarrow\infty$), many questions remain regarding its ground state, especially for finite $\noq$.  
Indeed, for typical instances, the value of the lowest energy is
known and efficiently classically computable~\cite{panchenko2013sherrington}.
This relative abundance of results and insights, combined with the fact that application problems are often conveniently mappable to weighted fully-connected Ising models~\cite{lucas2014ising}, made the SK an early target for an advanced benchmark of analog quantum and quantum-inspired solvers~\cite{mohseni2022ising}. 
Recently, it was shown that a classical  algorithm~\cite{montanari2021optimization} with high-probability returns a solution with energy at most a $(1-\epsilon)$ fraction of the optimal~\footnote{This claim relies on a conjecture related to replica symmetry breaking which is widely believed to be true though remains unproven.}, with a run-time that is polynomial in $\noq$ for fixed $\epsilon$. 
Complementary state-of-the-art classical approaches for the SK problem are based on semi-definite programming relaxations~\cite{bandeira2019computational}.

In the context of variational optimization, 
some exciting theoretical progress has been made for QAOA applied to the SK model in the $\noq\rightarrow\infty$ limit. 
In particular, \cite{farhi2022quantum,basso2021quantum} obtained analytic expressions for QAOA performance which indicates that 
QAOA requires only $p=11$ layers to outperform the standard 
semidefinite programming approach~\cite{montanari2016semidefinite}, and even down to $p=1$ when combined with a relax-and-round approach~\cite{dupont2023quantum2}.
It has been proven that for QAOA expectation value of the SK energy at optimal parameters concentrates for typical instances~\cite{farhi2022quantum}, indicating that parameter setting challenges may be significantly alleviated here. 

\paragraph*{Related experimental progress and challenges.}
A series of recent works~\cite{harrigan2021quantum,otterbach2017unsupervised,qiang2018large,abrams2019implementation,bengtsson2019quantum,willsch2020benchmarking,pagano2020quantum,niroula2022constrained,ebadi2022quantum,nguyen2023quantum,dupont2023quantum,shaydulin2023qaoa,pelofske2023high,pelofske2023scaling} has begun to explore the performance and limitations of standard QAOA and related approaches on today's real-world quantum hardware.

Until recently, the largest-scale prior study on the performance of QAOA on the fully connected Ising models
was carried out in Ref.~\cite{harrigan2021quantum} where instances up to size 
$\noq=17$  
and $p=3$ layers were compiled and run on the Google Sycamore quantum processor.
A benchmark study on how small-scale experimental results are deviating from numerics at optimized parameters was performed in \cite{tomesh2022supermarq} on IBM, IonQ, and AQT QPUs.
A constrained problem similar to SK, with real-valued coefficients out of financial market data, was tested in~\cite{baker2022wasserstein} on IBM, Rigetti, and IonQ QPUs. In \cite{dupont2023quantum}, we studied the same class of instances on the same hardware that we feature in this work, using a circuit ansatz consisting of a single-layer ($p=1$) truncated QAOA circuit, up to 72 qubits in a hybrid-recursive scheme.
Very recently, in \cite{sack2023largescale} the authors reach $p=2$ and 40 qubits on superconducting processors on 3-regular graphs, and in \cite{shaydulin2023evidence} a QAOA $p=1$ set of runs was performed for up to 18 qubits of a QCCD ion trap processor on a complex four-body Hamiltonian that can be compiled in a fully connected graph.
High-$p$ experimental studies of QAOA for up to $\noq=10$ have recently appeared in ionic QPUs~\cite{pelofske2023high} - but are still not operating at experimentally-optimized parameters, and they do show a performance rapidly converging to that of a random guess for sufficiently large depth.
In \cite{pelofske2023scaling}, large scale (with up to $n\approx 400$ qubits), short-depth (up to $\approx 500$ 2-qubit gates) QAOA circuits were implemented on IBM's quantum devices.

Unlike the approach of the above references, we do not focus on the comparison of the estimated expectation value of the cost Hamiltonian from their experiments to noiseless simulations. Instead, we explore the optimization-relevant performance of our solvers as stochastic optimization black boxes, in a regime where exact simulations are intractable.

We note that our motivation for benchmarking implementation of the fully connected graphs (specified by the Sherrington-Kirkpatrick from Eq.~\eqref{eq:H_C-SK}) was exactly that it is a relatively challenging task for limited-connectivity superconducting hardware.
Such dense problems are among the hardest benchmarks for testing the capabilities of state-of-the-art devices, and thus provide great test instances to showcase what can be currently achieved in large-scale experimental optimization.
See, e.g., overview of recent experimental demonstrations in Ref.~\cite{abbas2023quantum}.

\section{Hardware-Efficient ans\"atze}
\label{sec:ansatze}

Here we 
detail the new quantum ans\"atze we consider for experimental implementation, which are derived from previously proposed QAOA and QAMPA approaches. 
We consider linear-connectivity implementations utilizing SWAP networks, as well as new \textit{time-block} variants of QAOA/QAMPA that truncate the original circuit depth per layer.

\subsection{Linear connectivity implementation}

The description of both ans\"atze in 
Sec.~\ref{sec:ansatze_circuits} implicitly assumed that each 2-qubit gate can be implemented.
However, in practice, the connectivity of quantum devices is often restricted -- entangling gates can be physically implemented only between 
neighboring qubits. 
One method to circumvent this problem is to implement a SWAP network, i.e. a combination of SWAP gates that allows entangling qubits which cannot interact directly.
In this work, we implement a maximally-parallelized SWAP network architecture that requires only linear connectivity between qubits~\cite{hirata2011}.
Explicitly, each QAOA 
layer 
is implemented as 
\begin{align}\label{eq:qaoa_swap_network}
    \prod_{m=1}^{\noq}\prod_{i\in \mathcal{S}_{m}}\  ZZ^{(J)}_{i,i+1}\rbracket{\gamma} \ \text{SWAP}_{i,i+1} \ , 
\end{align}
where $\mathcal{S}_m$ denotes a set of indices specifying a linear chain of qubits (starting from index $1$ for $m$ odd and from index $2$ for $m$ even, see Fig~\ref{fig:paper_overview}) 
followed by an $X$-rotation gate applied to each qubit.
 
Similarly, we implement each QAMPA layer as
\begin{align}\label{eq:qampa_swap_network}
\prod_{m=1}^{\noq}\prod_{i\in \mathcal{S}_{m}}\  ZZ^{(J)}_{i,i+1}\rbracket{\gamma}\  XY_{i,i+1}\rbracket{\beta}\ \text{SWAP}_{i,i+1} \ . 
\end{align}
For simplicity, we impose a fixed gate ordering on the QAMPA ansatz (recall discussion in Section~\ref{sec:ansatze_circuits}).

Importantly, assuming CPHASE and XY rotations are in a native gate set (as is the case for Rigetti's chips), the above-described architecture for linear-chain SWAP network can be compiled using only $p\noq(\noq-1)$ physical two-qubit gates for $p$-layer ansatz.
To achieve this, we make use of the identity~\cite{larose2022mixer} (up to a global phase),
\begin{equation}\label{eq:identity_swaps}
\begin{split}
&ZZ_{i,j}^{(J)}\rbracket{\gamma}\  XY_{i,j}\rbracket{\beta}\ \text{SWAP} = \\ 
&=ZZ_{i,j}^{(J)}\rbracket{\gamma+\frac{\pi}{4J_{i,j}}} \ XY_{i,j}\rbracket{\beta+\frac{\pi}{4}} \ .
\end{split}
\end{equation}
Since the SWAPs are always accompanied by
ZZ rotations in the ansatz, 
the above allows compiling them via additional phase shifts for ZZ and XY gates, which reduces the number of required physical two-qubit gates for Eqs~\eqref{eq:qaoa_swap_network} and \eqref{eq:qampa_swap_network}. 
See Appendix~\ref{app:sec:compilation_details} for the explicit definitions of the above gates and 
compilation details.

\subsection{Time-block 
QAOA and QAMPA}\label{subsec:TBvariants}

Ans\"atze such as 
QAOA and QAMPA 
can 
require significant quantum resources 
even for a relatively small number of layers.
This motivates the search for ansatz circuits that are less resource-demanding while still allowing for high expressibility and performance.
A 
natural approach is to introduce parametrized layers that are shallower than those in a standard QAOA or QAMPA -- thus allowing for the same number of variational parameters controlling a circuit with smaller physical depth.
 
We propose a simple realization of such ans\"atze based on implementing only a part of the total Hamiltonian interactions in each layer, which we refer to 
as \textit{Time-Block} (TB) QAOA and QAMPA, respectively. 
While we consider fully-connected cost Hamiltonians without local fields (Eq.~\eqref{eq:fully_connected_hamiltonian}), our construction can be similarly applied to a variety of other problems.

Denote by $\mathcal{S}$  the set of indices of pairs of qubits, corresponding to the terms in $H_C$.
We 
divide the Hamiltonian interactions into multiple batches $\cbracket{\mathcal{S}_t}$ (with $\mathcal{S} = \bigcup_{t} \mathcal{S}_t)$) and implement each of them as a separate QAOA-like layer with a phase separator corresponding to
$H_C$ restricted to interactions in a given batch, i.e., $H_{\mathcal{S}_t} = \sum_{i,j\in\mathcal{S}_t}J_{i,j}Z_iZ_j$, followed by an independently parameterized mixing operator. 

For ansatz implemented via the SWAP network architecture described in the previous subsection, one can choose the TB division in such a way that it corresponds to dividing the original QAOA or QAMPA circuit (recall Eqns.~\eqref{eq:qaoa_swap_network} and \eqref{eq:qampa_swap_network}) into a series of 
sublayers, executed in temporally disjointed intervals. For instance, for QAMPA 
\begin{equation}\label{eq:idea_of_TB}
    \begin{split}
        \prod_{i,j\in \mathcal{S}}\tilde{u}_{i,j}
\longrightarrow \prod_{t=1}^{s} \left[\prod_{i,j\in \mathcal{S}_t}\tilde{u}_{i,j}\right] \ ,
    \end{split}
\end{equation}
where for simplicity we've dropped the dependence on the variational parameters.

Comparing Eq.~\eqref{eq:idea_of_TB} with Eqs.~\eqref{eq:qaoa_swap_network}, \eqref{eq:qampa_swap_network} suggests a choice of the division of Hamiltonian interactions that allows minimizing the physical depth of the implemented SWAP network.
Namely, we choose a division that cuts the full original ansatz circuit into parts that consist of $k$ parallel executions of two-qubit gates in a checkerboard pattern.
In each layer, this corresponds to implementing $\approx k\noq$ interactions from the total $\binom{\noq}2\approx\noq^2/2$ present in the Hamiltonian.

This construction is illustrated in Fig~\ref{fig:paper_overview}
for two choices of $k$ ($k$=1 and $k$=2).
In the rest of the paper, we will focus on this particular type of time-block ansatz, and we will refer to it as TB $k$-QAOA ($k$-QAMPA) following the $k$ letter with its chosen value (see Appendix~\ref{app:expDetails} for explicit construction).
Importantly, the two-qubit gates needed for ansatz construction are compiled using identity from Eq.~\eqref{eq:identity_swaps},
allowing implementation of TB $k$-QAOA/QAMPA ansatz with $p$ layers using only $\mathcal{O}\rbracket{pkn}$ many 2-qubit gates, assuming native availability of $XY$ and CPHASE~\footnote{We note that $ZZ\cdot XY$ interactions could also be implemented using a single application of parametric fSim gate (and 1-qubit rotations) of Google's Sycamore processors \cite{arute2019quantum}, or using 3 applications of M{\o}lmer–S{\o}rensen (MS) gate (and 1-qubit rotations) commonly used in ion-traps architectures \cite{Maslov2017ions}. 
Our ans{\"a}tze are thus suitable for efficient implementation on QPUs different than those benchmarked in this work.}.

We 
emphasize that TB ansatz is a more structured generalization of the idea to allow each qubit gate to have its own different parameter, a setting for which the quantum circuit resembles somewhat a quantum neural network with weights to be trained~\cite{farhi2017quantum,herrman2021multi,chalupnik2022augmenting}. 
Our ansatz design seeks to maintain a relationship to the structure of the cost function, and a thematic connection to the phenomenology of adiabatic quantum computing, where the angle parameters vary as a continuous function of time.

\begin{rem}\label{rem:tb_vs_standard}
TB $k$-QAOA (TB $k$-QAMPA) recovers a single layer of the underlying QAOA (QAMPA) ansatz
by choosing the local block size $k=\noq$ and the number of layers $p=1$, or by $k=1$ and $p=\noq$ with all angles set to the same value (in the case of QAOA, this also requires setting angles for single-qubit mixers in all layers besides the last to 0).
In general, if $k$ divides $\noq$, then $p=\frac{\noq}{k}$ layers can be implemented to recover a single layer of the standard ansatz. 
\end{rem}

We note that the Time-Block strategy of dividing the cost Hamiltonian into independently parametrized batches could be implemented with any ansatz that exploits the Hamiltonian structure in its construction. 
More generally, the variant of the over-parametrization strategy we applied to linear-chain SWAP networks (TB $k$-QAOA and $k$-QAMPA) could be applied to any hardware-efficient ansatz.
Our motivation to build specifically upon QAOA and QAMPA was the following.
On the one hand, the QAOA is a canonical example of quantum approximate optimization ansatz.
As such, it constitutes a standard baseline to test against, as is usually done in the literature (see, e.g., overview paper \cite{abbas2023quantum}).
On the other hand, the motivation behind the choice of QAMPA was mainly practical -- the $XY$ mixer used in QAMPA is a native Aspen-M-3 gate (see Appendix~\ref{app:sec:compilation_details}), hence its implementation is efficient "for free" in our experimental setup.

\subsection{Ansatz Design Optimization}\label{sec:ansatze_permutations}

Inspection of Fig.~\ref{fig:paper_overview} (see also Eqs.~\eqref{eq:qaoa_TB_swap_network} and \eqref{eq:qampa_TB_swap_network} in Appendix~\ref{app:sec:compilation_details})
 makes it clear that in the construction of the TB $k$-QAOA/QAMPA ansatz circuit, the SWAP network will determine which two-body interactions should be included in the local blocks (layers) of the circuit. and in what order.
Indeed, each choice of the (ordered) sets of interactions corresponds to a different ansatz for both QAMPA and QAOA (however, see 
Remark~\ref{rem:TB2} below), and can thus lead to a different performance in variational optimization.
For simplicity, 
we refer to the setup specifying which interactions are appearing and in what order as a \emph{gate ordering} (or GO).

In the considered ansatz, the gate ordering can be viewed as an additional categorical parameter 
of the circuit and thus it can be optimized over.
Orderings are fully identified by the initial variable assignments to the linear register of qubits, hence their number grows factorially with $\noq$, which makes exhaustive optimization 
intractable even for small systems. 
To circumvent this difficulty we propose choosing a number of random orderings and optimizing over that 
set (as a categorical variable), see also Remark~\ref{rem:GOs} below.
In Appendix~\ref{app:sec:results_angles_permutations}, we experimentally observe that the choice of gate ordering can have significant effects on 
algorithm performance, and thus it is 
indeed advisable to consider different orderings for the quantum devices we used.
We note that one could attempt to optimize ans\"atze by following physical guidance in the choice of orderings -- for example, 
using experiments or noise models that incorporate cross-talk for a given problem, a strategy akin to noise-aware compilation~\cite{murali2019noise}.  
We provide some additional discussion on the effect of gate ordering in Appendix~\ref{app:sec:results_angles_permutations}.

\begin{rem}\label{rem:GOs}
    We note that \emph{a priori}, our strategy of choosing Gate Orderings at random might seem to be bound to fail due to a factorially growing number of all possible orderings.
    However, it turns out that in practice, for experiments on as much as $\noq=50$, it suffices to optimize over only $\approx 10$ GOs to achieve visible performance gains, as we discuss in Section~\ref{sec:results}.
    In particular, in Fig.~\ref{fig:best_results}, the comparison between dashed lines (performance averaged over $10$ random GOs) and solid lines (performance of the best out of $10$ random GOs) for random angles optimizer shows an improvement by a factor of $\approx 2-3$. 
    Similar performance gains are obtained for smaller system sizes in Appendix~\ref{app:sec:results_small_n}. 
\end{rem}

\begin{rem} \label{rem:TB2}
For TB $k$-QAOA 
with $k=\noq$, when the standard QAOA ansatz is recovered, different orderings of gates within the time block correspond to the same logical circuit (as all of the gates commute), which is not the case for arbitrary $k<\noq$.
However, even in this case, implementing gates in a different order might lead to a 
noticeable performance difference in experiments, due to the asymmetric effects of noise.
In general, adaptively permuting gates within a layer can be viewed as an error-mitigation method~\cite{hashim2022optimized}.
\end{rem}

\begin{rem}\label{rem:adaptQAOA}
We note that our Time-Block construction and optimization of gate orderings is conceptually related to the recently proposed ADAPT-QAOA algorithm \cite{Zhu2022AdaptQAOAOriginal}.
In ADAPT-QAOA, in the course of optimization, the mixer operator is allowed to be modified, allowing more flexibility and adaptivity.
In the case of Time-Block ansatz, different choices of Hamiltonian interactions batching, parameter $k$, as well as the gate orderings, can be interpreted as implementing a different phase separation (driver) Hamiltonian (see, e.g., recent related approaches in Ref.~\cite{wilkie2024qaoa,maciejewski2024ndar}).
Since GOs are chosen by the optimizer, this variant of optimization can be viewed as adaptively improving the phase separator.
\end{rem}

\section{Experimental methodology}\label{sec:methodology}

Here we provide the main details of our experimental approach. Some additional technical details are deferred to Appendix ~\ref{app:expDetails}.

\subsection{Assessing optimization quality}\label{sec:methodology_quality}

To quantify the quality of a classical bitstring solution $x$ with cost $C(x)=\langle x|H_C|x\rangle$, 
we utilize the \textit{approximation ratio} $r_\costfunction$ defined as 
\begin{align}\label{eq:ar_definition}
    r_{\costfunction} = \frac{\costfunction(x)}{\costfunction_{\min}}\,\leq 1\ ,
\end{align}
where 
$\costfunction_{\min}$ is the
minimum cost. 
In particular,
$r_{\costfunction}$ obtains its extremal value~$1$ minimization returned $\costfunction_{\min}$, when an optimal solution is returned.
Note that with this definition $r_\costfunction$ becomes negative for sufficiently poor (opposite sign) solutions.
 We will experimentally assess the performance of our quantum 
ansatz using empirical estimates of $r_\costfunction$ obtained across multiple problem instances.
To find $\costfunction_{\min}$ classically, we use a chordal branch-and-bound method introduced in \cite{Baccari2020} in the context of benchmarking quantum annealers. 

In order to compute estimators of the approximation ratio $\langle r_C \rangle$, each experiment entails measuring the energy of the $H_{\costfunction}$ on a variational state $s$ times (``shots").
To guide the variational optimization, typically the standard estimator for an expectation value is used
\begin{equation}\label{eq:ar_singleshot}
    \langle C \rangle = \frac{1}{s}\sum_{k=1}^{s}C_k \ ,
\end{equation}
where $C_k\equiv C(x_k)$ is the energy value of $k$th sample $x_k$~\footnote{Note that this expectation value estimates the result from a \emph{single shot} execution of the circuit.}.

In Appendix~\ref{app:results_tails}, we also consider estimators constructed from low-energy tails of empirical distributions.
The goal is to study whether a given distribution is more likely (than a random baseline) to reach high-quality solutions.
Having in mind that in binary optimization one is interested in \emph{single} best solution (as opposed to expected values typically used to guide the optimization), we believe that this type of analysis, coupled to the quantification of the cost of finding good parameters, is especially practically relevant and we intend to extend it in future work~\cite{windowsticker}.

\subsection{Uniform random sampling}\label{sec:methodology_baseline}

We compare the outcomes of our experiments with the results of a random-guessing algorithm consisting of uniform random bitstring sampling.
Random bitstring sampling is ideally realized on a quantum device in a strong noise regime when one effectively samples from a maximally mixed state, and the resulting outcomes appear uniformly random. Moreover, it is the 
simplest, most generic classical heuristics to compare against.
Observe that for uniform random sampling, the expectation value of the cost Hamiltonian in 
\eqref{eq:H_C-SK} is easily seen to be $0$,
and thus the approximation ratio is $\langle r_{\costfunction}^{\text{random}} \rangle = 0$ (in the limit of infinite number of samples).

\begin{figure*}[!t]
\includegraphics[width=1.0\textwidth]{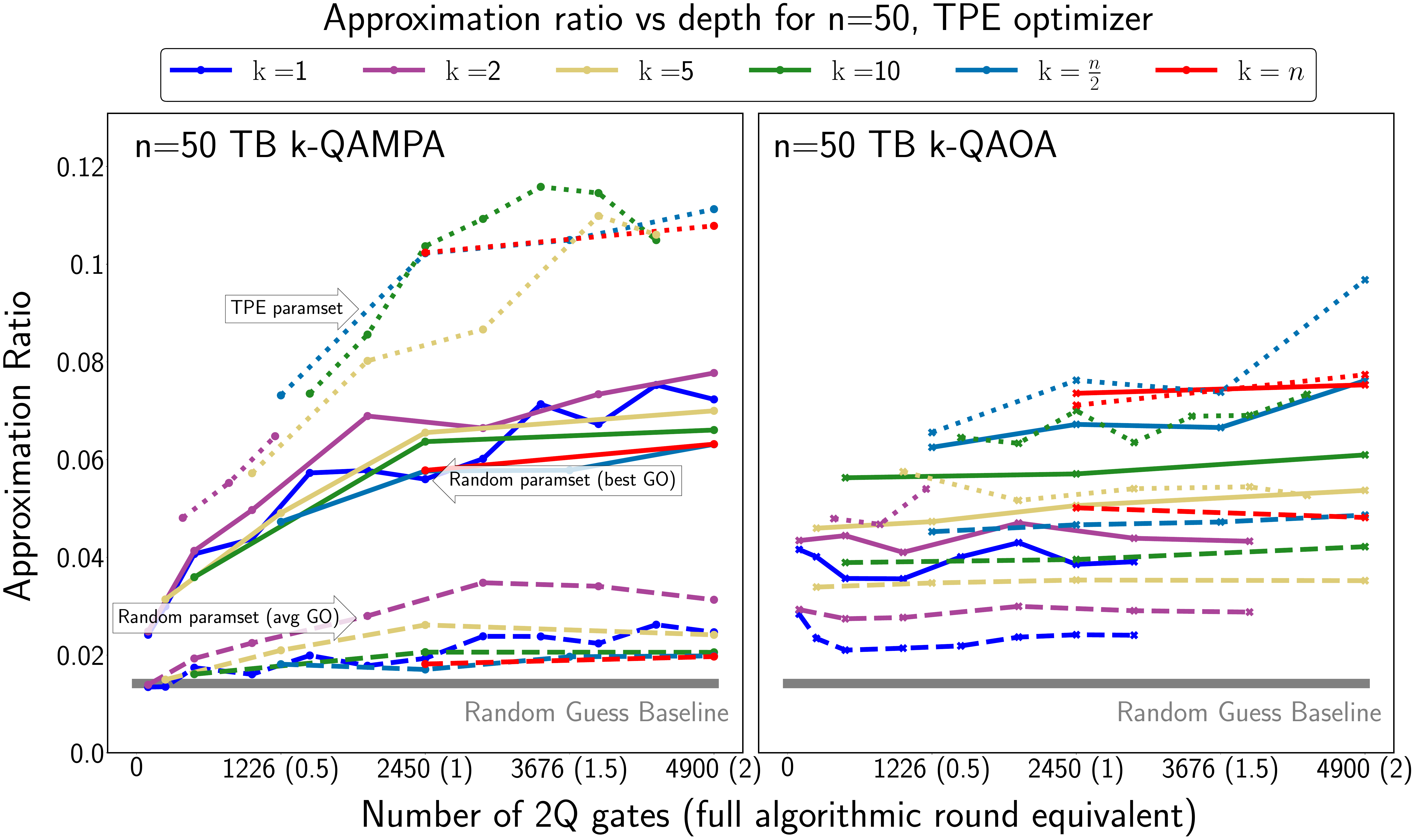} 
\caption{Experimental results of $\noq=50$ QAMPA (left) and QAOA (right) implemented using ans\"atze introduced in this work. 
The vertical axis corresponds to the approximation ratio estimated using \eqref{eq:ar_singleshot}).
The horizontal axis shows the depth of the circuit, i.e., physical no. of 2-qubit gates (the fraction of standard QAMPA/QAOA ansatz logical layers, see Remark~\ref{rem:tb_vs_standard} and discussion in the text). 
Different colors correspond to different ansatz circuit architectures (see Section~\ref{sec:ansatze_circuits}).
Our ans\"atze admit optimization over the ordering of circuit gates.
The solid lines correspond to optimized gate ordering (``GO'' in the plot), chosen as the best out of 10 randomly generated, while dashed lines to the average gate ordering (see Section~\ref{sec:ansatze_permutations}), both using the random optimizer. 
The dotted lines correspond to optimization over both angles and gate ordering using adaptive Tree of Parzen Estimators (TPE) optimizer \cite{hyperopt2013}.
  \label{fig:best_results}}
\end{figure*}

\subsection{Experimental implementation}\label{sec:methodology_setup}

\begin{table}[t!]
\begin{tabular}{|c|c|}
\hline
\textbf{Parameter}                     & \textbf{Values in experiments} \\ \hline
\textit{number of qubits (n)}          & \{12, 16, 20, 30, 40, 50\}     \\ \hline
\textit{size of the local block (k)}   & \{1, 2, 5, 10, \noq/2, n\}        \\ \hline
\textit{number of layers (p)}          & $\leq 2\noq/k$            \\ \hline
\textit{number of trials}          & $\cbracket{1000, 800}$                           \\ \hline
\textit{number of shots per trial (s)} & 1000                           \\ \hline
\textit{number of gate orderings}    & $\cbracket{10, 50}$       \\ \hline
 \textit{parameter setting strategy}                     & random or TPE   \\ \hline
\textit{Hamiltonian instance}          & {[}1,10{]}                     \\ \hline
\end{tabular}
\caption{Description of experimental parameters, together with exemplary implemented values.}\label{tab:exp_summary}
\end{table}

We report the implementation of a multiqubit variational optimization with time-block $k$-QAOA and $k$-QAMPA ans\"atze on subsystems of Rigetti's $79$-qubit superconducting quantum processor Aspen-M-3.

The experiments were performed for multiple system sizes $\noq \in \cbracket{12, 20, 30, 40, 50}$. 
For each system size, $\noH$ random instances of the Sherrington-Kirkpatrick model without local fields (see Eq.~\eqref{eq:fully_connected_hamiltonian}) were implemented (see also Appendix~\ref{app:sec:hamiltonians_gs}).
Each instance was implemented with TB $k$-QAOA and $k$-QAMPA ans\"atze with varying block sizes $k$, and for each $k$ multiple depths $p$ were implemented. 

In our experiments, we choose local block sizes $k\in\cbracket{1,2,5,10,\noq/2,\noq}$, and for each $k$ we implement circuits with multiple depths $p \in \sbracket{1,\dots,2 \noq / k}$.
We express the physical circuit depth of each circuit as a fraction of the number of gates one would need in order to implement standard QAOA/QAMPA (recall Remark~\ref{rem:tb_vs_standard}).

In most experiments, for each triple $\rbracket{\noq, p,k}$,
we implemented $100$ random sets of angles and for each set performed computational-basis measurement on the variational state for a number of $s=1000$ shots. 
Each set of random angles was implemented $10$ times, each with different, random ansatz gate ordering (recall \ref{sec:ansatze_circuits}), for a total of $t=100\times10=1000$ \emph{trials}.
This corresponds to the simplest, non-adaptive random parameter setting strategy.
For $\noq = 50$, we further implemented an adaptive parameter setting strategy using a black-box optimizer based on Tree of Parzen Estimators (TPE)~\cite{Bergstra2011,hyperopt2013} as implemented by the package \texttt{optuna} \cite{optuna2019} (see Appendix~\ref{app:sec:results_angles_permutations} for details), evaluated $\approx 800$ times~\footnote{We did not reach $t=1000$ trials due to constraints on available QPU time.}. 
We allowed the TPE parameter optimizer to search also over a categorical variable that controls gate ordering, chosen as one of 50, randomly generated.

The summary of parameters specifying each implemented experiment is presented in Table~\ref{tab:exp_summary}.

\section{Experimental results}\label{sec:results}

In this section, we present the results of the experiments.
Additional analysis is presented in Appendix~\ref{app:additional_results}, including a detailed analysis of the tails of obtained energy distributions in Appendix~\ref{app:results_tails}.

\subsection{Quality of solution vs algorithmic depth}\label{sec:results_quality}

We study the dependence between the average (over $\noH$ random SK-model instances) approximation ratio, as a function of the \emph{physical depth} of the circuit for $\noq=50$ in Fig.~\ref{fig:best_results}.
This depth is expressed i) as a number of physical 2-qubit gates, and ii) as a fraction of the algorithmic depth (number of layers, usually indicated by $p$) that would be needed to implement the standard QAOA/QAMPA, as presented in the original papers ~\cite{farhi2014quantum,larose2022mixer}.

The fraction ii) is $\approx pk/\noq$ (this is because $\approx \frac{n}{k}$ layers of TB $k$-QAMPA/QAOA are needed to recover a single layer of standard ansatz, see 
Remark~\ref{rem:tb_vs_standard}). 
For low $k$, the TB ansatz with physical depth similar to the standard QAOA/QAMPA is specified by a significantly higher number of variational parameters -- and thus, in general, it is harder to optimize. 

The thick lines in Fig.~\ref{fig:best_results} show the results for $\noq=50$ (results for smaller $\noq$ are reported in the Appendix), comparing the approximation ratio estimator (constructed from $1000$ samples) obtained utilizing the best gate ordering and the best set of angles (i.e., the best values out of $1000$ trials) obtained using the random parameter setting strategy.
To make the comparison fair, for the random guess \emph{baseline} we report estimation for the best obtained $r_C$ estimator (constructed from $1000$ samples) from $1000$ random sampling repetitions (same as the total number of trials implemented for each $k$ and $p$)~\footnote{Note that this is not common practice - usually experimental results at best parameters are compared to the single-shot random guess estimator with disregard to the additional resources needed to find the best parameters.}.
In most cases, we observe that, on average, the QPU solvers outperform the random sampling baseline. 
Remarkably, for QAMPA we observe a roughly monotonic trend where performance increases with physical circuit depth (note that the parameter setting optimization is independent for each $p$) before reaching a plateau.
We note that for high $p$ the experiments involve implementing thousands of single and two-qubit gates ($pk/n=2$ corresponds to $\approx5000$ gates), yet we are able to recover visible monotonicity and beat random baseline for almost all data points. 
On the other hand, for QAOA we do not observe an appreciable growth of average performance as we increase the number of TB layers of the circuit. 
Moreover, our results indicate that the QAOA performance increases with $k$ at constant circuit depth. 
In particular, the standard ($k=n$) QAOA ansatz achieves the best absolute performance, indicating that the TB parametrization, while not hurting in principle, is not manifestly a useful variant of QAOA in practice.

Dashed lines in Fig.~\ref{fig:best_results} report on the $\langle r_C \rangle$ for a random gate ordering, averaging the best performing among 100 angle trials over 10 gate orderings resulting from random initial qubit assignments. While the qualitative trend of performance vs depth is preserved without gate orderings optimization, the quantitative improvement is striking. More results and discussion on the comparative performance of optimizing the gate orderings vs the angle parameters can be found in Appendix \ref{app:sec:results_angles_permutations}. 
We should note that this study that decouples the performance sensitivity of tuning the GOs versus the angles has been possible mostly because of our choice to benchmark the random parameter setting strategy.
Here the trials are identified as the Cartesian product of random vectors of parameters. 
For more elaborate, adaptive parameter setting strategies, a more refined analysis would be required~\cite{windowsticker}.

For a subset of $k$s and $p$s (namely $k=\{2, 10, n/2, n\}$ up to $2$ equivalent depth of QAOA/QAMPA), we also implemented an adaptive optimization method known as the Tree of Parzen Estimators (TPE). 
The optimizer was allowed to optimize over categorical variables controlling the ansatz gate ordering (out of $50$ random orderings). The choices of these meta-parameters have not been optimized for this work, the study is meant to provide a preliminary comparative analysis of the improvement we could expect by adaptive parameter setting strategies.

The results of TPE optimizations are presented as dotted lines in Fig.~\ref{fig:best_results}.
We observe that for both QAMPA and QAOA the TPE optimization delivers an advantage over the simple strategy of random parameter setting.
It is notable to observe that optimization over gate orderings provides an advantage for both ans\"atze, with QAMPA performing very poorly without it (see Appendix~\ref{app:sec:results_angles_permutations} for discussion of this effect).

\subsection{Performance scaling with problem size}\label{sec:results_scaling}
\begin{figure*}[t]
\begin{center}
\includegraphics[width=1.0\textwidth]{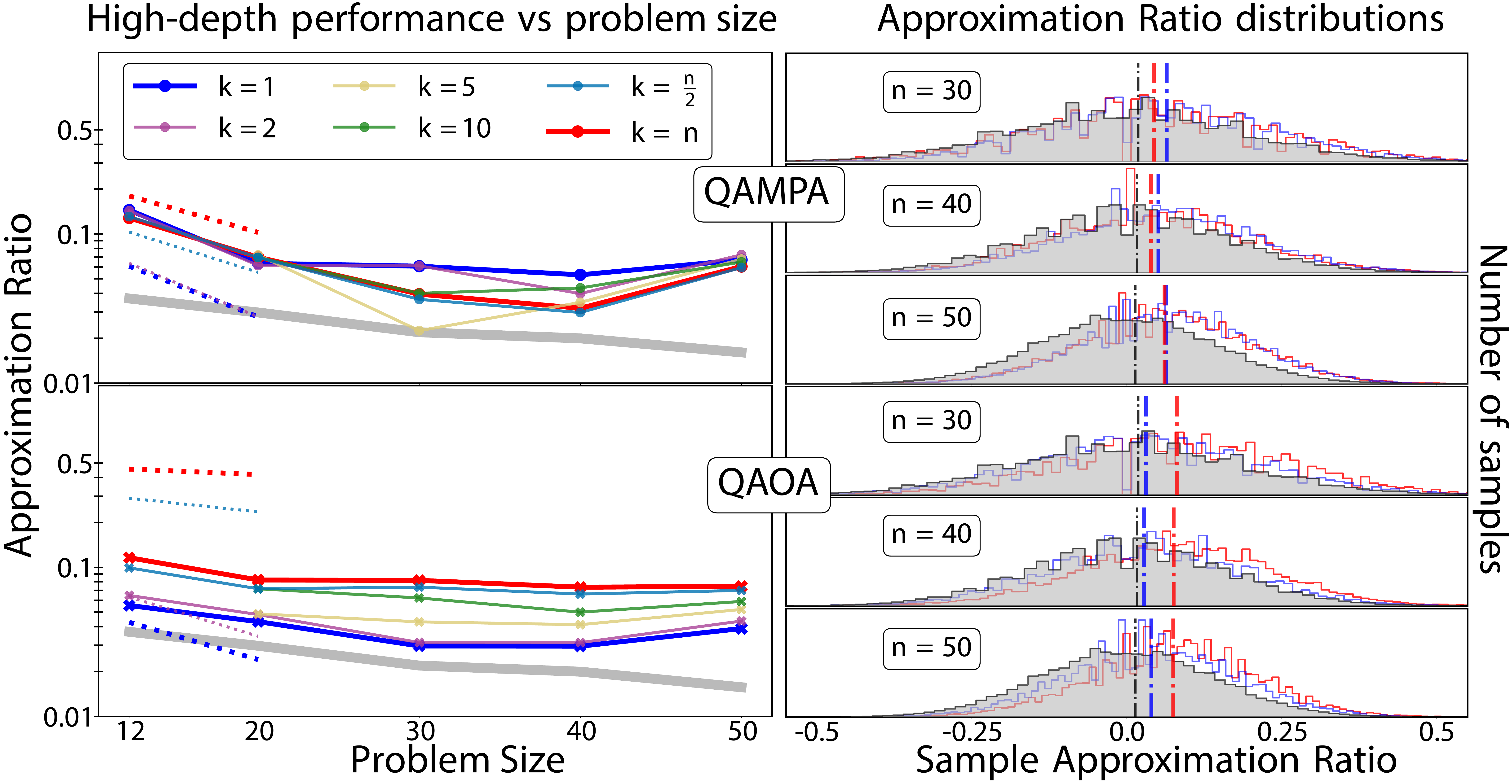}
  \caption{
  The left-hand side plot shows an average (over $\noH$ random Hamiltonian instances) estimated approximation ratio as a function of system size.
  For each $n$ and $k$, each datapoint is mean over depths corresponding to range $\frac{pk}{n} \in \sbracket{1,2}$, i.e., between depth 1 and 2 of a standard ansatz.
  The dotted lines correspond to the wavefunction simulator.
  The simulator is run on the same parameters' sets as corresponding experiments (though optimal values may differ). 
 Similarly, the other specifications, including the number of trials and samples, are the same as for experiments.
  The right-hand side plot shows energy outcome distributions for  $k=1,\ n$ with depths corresponding to $\frac{pk}{n}\in\sbracket{1,2}$, as compared to random bitstring sampling.
The dashed vertical lines correspond to the averages reported in the left-hand side plot.
}\label{fig:scaling_with_n}
 \end{center}
\end{figure*}

As system size grows, combinatorial problems generally become harder to solve,
and so investigating the scaling of the 
performance 
with system size is of high practical importance.
We plot the results using the random parameter setting strategy in the left-hand side of Fig.~\ref{fig:scaling_with_n} as a function of the number of qubits $\noq$ for each $k$.
In the right-hand side, we plot the approximation ratio distribution of measurement outcomes of our studied solvers for $k=1$ and $k=\noq$, compared against the baseline obtained from uniform random sampling.

For fixed $k$ and $\noq$, the data corresponds to the average performance obtained for experiments at a full algorithmic round equivalent depth between 1 and 2 (inclusive), ranked via the value of the mean approximation ratio, in order to capture the plateau value of performance observed in Figure \ref{fig:best_results}.
In both cases, the experimental distributions from QPU runs are visibly centered to the right (toward higher values of the approximation ratio) w.r.t. the random baseline~\footnote{We should note that the fact that $\noq = 50$ seems to perform better than $\noq = 30$ and $40$ is likely due both to statistical fluctuations as well as the fact that the sublattice selection is different.}. However, as discussed in Appendix~\ref{app:results_tails}, the advantage obtained comparing the mean of the distribution is less significant towards the tails.

More detailed analysis of experiments performed on smaller system sizes is presented in Appendix~\ref{app:sec:results_small_n}. 
We also present a restricted analysis of the effects of choosing the Hamiltonian mapping that can be beneficial or adversarial due to particular noise characteristics in a device.
To this aim, we follow methods from recent work ~\cite{maciejewski2024ndar}, where a noise-aware quantum approximate optimization method was proposed by some of the authors of the present paper. 

We note that the data presented in Figure~\ref{fig:best_results} corresponds to \emph{mean} approximation ratio (estimator constructed from $1000$ samples of the best trial).
In practice, the actual solution to the binary optimization problem is a single, \emph{best} bitstring. 
The best-found distributions can be seen on the right-hand side of Figure~\ref{fig:scaling_with_n}, where the tails of the approximation ratio distributions have appreciably higher values than the means presented in Figure~\ref{fig:best_results}.

It is clear that the overall performance can not, at its current capabilities, compete with state-of-the-art classical solvers (see, e.g., Refs.~\cite{arora1995polynomial, fernandez1996max, fernandez2000polynomial, farhi2022quantum, montanari2021optimization, abbas2023quantum}).
However, our experiments showcase that the quantum device, after suitable parameter optimization, is capable of producing solution distributions with moderately better properties than the repeated random guessing baseline (for as much as $50$ qubits and thousands of 2-qubit gates).
This is certainly a necessary step on the path toward potential quantum utility \cite{herrmann2023quantum}.

\section{Discussion}\label{sec:discussion}

In this paper, we introduced novel ansatz design techniques for quantum optimization, with the objective to maximize the potential of NISQ devices to solve challenging highly-connected optimization problems.
The Time-Block $k$-QAMPA/QAOA ansatz divides the original Hamiltonian into batches that correspond to parameterized ansatz layers consisting of $k$ sub-circuits compiled on linear chains of qubits.
Due to the use of an efficient SWAP network architecture and compilation scheme supporting native gates on devices such as Aspen-M-3, our $p$-layer ansatz can be implemented using $\mathcal{O}\rbracket{pkn}$ 2-qubit gates.

We tested our approach in experiments on up to $50$-qubit subsystems of Rigetti's QPU for random instances of the Sherrington-Kirpatrick model.
In all experiments, including circuits involving $\approx 5000$ two-qubit gates, we observed the experimental optimization to perform better than the uniform bitstring sampling comparative baseline.
For QAMPA, there is also a clear benefit in the Time Block parametrization and in the constructive increase of $p$ and corresponding circuit depth. For all cases studied, we observe a substantial increase in performance by introducing a novel ansatz tuning knob: the gate orderings. Among all our tests, for the approximation ratio we obtain approximately $\simeq 0.1$, to be compared with a random guess baseline of $\simeq 0.015$. 

This work points to various followup research directions across multiple dimensions of theory, engineering, experimental design, and data-analysis.

\paragraph*{Noise and phenomenological modeling.} The presented experiments clearly operate the QPU in a regime where noise dominates the quantum state dynamics. It will be interesting to gain more insights on the magnitude, character, and effects of noise, for instance, by running tests that could witness the presence of coherence and entanglement as a function of circuit length~\cite{alam2022practical}. More generally, together with smaller-scale microscopic simulations, we believe it is interesting to study to what extent our results can be described by a dissipative phenomenological model relaxation of our TB parametrization scheme, both for QAOA and QAMPA, considering that each individually parametrized time-block has a very similar structure which borrows some resemblance from quantum annealing schedules with XY interactions~\cite{hen2016quantum} and from digital-analog optimization approaches~\cite{headley2022approximating}.

In Appendix \ref{app:sec:encodings}, we offer a preliminary exploration of strategies to mitigate some of the limitations imposed by hardware noise, drawing on techniques from Ref. \cite{maciejewski2024ndar}.
It remains an interesting future research direction to understand the interplay between noise characteristics (such as the $T_1$ and $T_2$ times of the qubits), chosen ansatz, and the parameter setting strategy required to achieve performant optimization.

\paragraph*{Theory and benchmarking of ansatz parameter setting.} We empirically demonstrated and studied the tunability of our solver by exploring and adjusting the TB angles as well as the gate orderings - however, this experimental evidence is in need of further analysis from the theoretical standpoint. In particular, it would be interesting to study concentration effects in parameter space~\cite{akshay2021parameter} for the idealized noiseless TB k-QAOA k-QAMPA algorithms, understanding rigorously why QAMPA is deriving benefits from the TB parametrization while QAOA is not, as well as to try to characterize the energy landscapes, quantifying effects such as presence of plentiful local minima \cite{wierichs2020avoiding,you2021exponentially,anschuetz2022quantum} and barren plateaus \cite{mcclean2018barren,cerezo2021cost,wang2021noise}, and to what degree modified quantum ans\"atze are able to alleviate these difficulties.

Our benchmarking analysis of parameter setting is also just a first step towards understanding what is achievable in practice in terms of performance. We have preliminary evidence from the adaptive TPE optimization that moving beyond random parameter setting exploration is a profitable endeavor, at least for the moderate size of the time blocks $k$. Additional strategies such as doubly stochastic gradient descents~\cite{Sweke2020stochasticgradient} or designed heuristics from analytical insights~\cite{sureshbabu2023parameter} are of interest. Besides the tuning of the \texttt{hyperopt} TPE meta-heuristics parameters, other knobs are to be explored such as classical spin-reversal symmetry transforms such as the ones used to optimize the performance of quantum annealers~\cite{shaydulin2021classical} (see also discussion in Appendix~\ref{app:sec:results_small_n} and recent related work \cite{maciejewski2024ndar}), or spatial permutation symmetries that lead to more efficient circuit constructions \cite{sauvage2024building}.
In our experiments, we discovered that optimizing over gate orderings can provide significant further performance gains, even when only a small number of them are taken into account.
This finding opens up an avenue for investigating the interplay between gate ordering optimization and cross-talk effects in gate noise, a phenomenon commonly observed in superconducting QPUs \cite{Zhao2022crosstalk}.
One potential application of this insight could be the development of heuristics that select gate orderings based on device-specific noise characteristics.

Applying Time-Block decomposition to circuit ans\"atze different than standard QAOA and QAMPA is another promising research direction.
For example, the approach of ADAPT-QAOA \cite{Zhu2022AdaptQAOAOriginal} to adaptively modify mixer Hamiltonians can be viewed as complementary to ours (see Remark~\ref{rem:adaptQAOA}), and could lead to improved results.

\paragraph*{Real-world performance evaluation in hybrid iterative algorithmic setting.} Our optimized expectation values are relatively superior to previously reported results on similar quantum setups, but they fall short of being exciting from a practical quantum advantage perspective. More precisely, advanced classical heuristics can likely solve SK problem instances at $\noq\simeq 100$ within milliseconds~\cite{mohseni2022ising}.
To make an order of magnitude comparison, in our experiments, we can extrapolate from the extremal statistics of the data presented in Fig.~\ref{fig:scaling_with_n} and in appendix \ref{app:results_tails} that we can achieve an optimization quality of about $\langle r_C \rangle \simeq 0.36$ considering the best bitstring from about 40 runs of the solvers for $\noq = 50$ (for an estimated total duration of approx 30 milliseconds including overhead times) at optimal parameters~\footnote{The estimate is based on measured wallclock QPU time and 40 runs correspond to what is estimated to be required to sample the best 2.5\% percentile, i.e. the average $\langle r_C \rangle$ from the top 50/1000 shots at optimal parameters.}. 

In future follow-up studies, it would be interesting to consider more challenging $J_{i,j}$ distributions (e.g. those described in \cite{perera2020chook}) and evaluate the time-to-quality metrics while including the parameter setting costs in the analysis~\cite{windowsticker, weidenfeller2022scaling}. However, the most important task will be to  dramatically boost the performance, beyond beating random guessing and toward outperforming classical heuristics such as simulated annealing. 
A promising
path to potentially achieving these goals is to employ alternative hybrid quantum-classical protocols, such as the iterative problem reduction approaches of
~\cite{bravyi2020obstacles, dupont2023quantum} (see also \cite{brady2023iterative}), or the recently proposed quantum relax-and-round solver~\cite{dupont2023quantum2}. Indeed, in \cite{dupont2023quantum} we observed an $\langle r_C \rangle \simeq 0.84$  for $\noq = 72$ on a shallow circuit algorithm that bears resemblance to the TB $k$-QAOA ($k=3$) algorithm for $p=1$ time block execution. It is plausible that by executing the best performing, optimized TB $k$-QAOA solver in a similar recursive scheme at high depth we could achieve better approximation ratios. Moreover, these iterative decompositions are susceptible to the use of standard ``weak" error-mitigation techniques  \cite{quek2023exponentially} to boost the fidelity of observables. 
Applying such methods to recursive optimization variants would 
lead to strong-error mitigation (i.e., access to samples from noiseless probability distributions).

\section{Conclusions}\label{sec:conclusion}

In conclusion, this study provides a benchmark of the state of NISQ optimization at the edge of what is possible in state-of-the-art superconducting quantum processors. In a departure from other works, our research emphasis has been on quantifying the bare performance of the solver relevant for optimization towards regimes of classical intractability~\cite{dupont2022calibrating,dupont2022entanglement} 
and identifying design and fine-tuning elements that are promising in view of a future hybrid solver design that might deliver quantum advantage.
We should note that the analysis of experimental results at this scale and with this level of co-design sophistication is an expensive endeavor utilizing still scarce resources: we estimate to have used about 520 hours of dedicated QPU access to perform all runs that aided this research project. This highlights one of the challenges faced in research and development today for the generation of sufficient data for proper benchmarking of quantum heuristics at the limit of what is possible with near-term, non-fault tolerant technology.

\section*{Acknowledgments}
This work was 
supported by the Defense Advanced Research Projects Agency
(DARPA) under Agreement No. HR00112090058 and IAA
8839, Annex 114. Authors from USRA also acknowledge
support under NASA Academic Mission Services under contract No. NNA16BD14C. B. H. acknowledges support from USRA Feynman Quantum Academy internship program.

\bibliography{bibliography}

\appendix\label{appendix}
\onecolumngrid

\section{Additional experimental details} \label{app:expDetails}

\subsection{Compilation}\label{app:sec:compilation_details}

For $p$ layers, the $k$-QAOA/$k$-QAMPA Time-Block ansatz circuit discussed in the main text (recall the circuits in Fig.~\ref{fig:paper_overview}~\footnote{Note that the first set of SWAPs after the Hadamards initialization is superfluous and can be removed in practice.}) is given by
\begin{equation}\label{eq:qaoa_TB_swap_network}  
\rbracket{\text{TB }k\text{-QAOA}}\ \prod_{t=1}^{p}\left[    \prod_{l=1}^{k}\prod_{i\in \mathcal{S}_{l}}\  u_{i,i+1}(\gamma_t)\right]U_{B}\rbracket{\beta_t}, 
\end{equation}
\begin{equation}\label{eq:qampa_TB_swap_network} 
\rbracket{\text{TB }k\text{-QAMPA}}\ \prod_{t=1}^{p} \left[   \prod_{l=1}^{k}\prod_{i\in \mathcal{S}_{l}}\  \tilde{u}_{i,i+1}(\gamma_t,\beta_t)\right] \ ,
\end{equation}
where, explicitly, $\mathcal{S}_{l} = \cbracket{1,3,\dots,\noq^{\rbracket{\text{l odd}}}}$ for $l$ odd, and  $\mathcal{S}_{l} = \cbracket{2,4,\dots,\noq^{\rbracket{\text{even}}}}$ for $l$ even. 
The last index depends on the parity of $\noq$.
Namely, $\noq^{\rbracket{\text{l odd}}} = \noq-1$ for $\noq$ odd and $\noq^{\rbracket{\text{l odd}}} = \noq-2$ for $\noq$ even, while  $\noq^{\rbracket{\text{even}}} = \noq-2$ for $\noq$ odd and $\noq^{\rbracket{\text{l odd}}} = \noq-1$ for $\noq$ even.
The local unitaries we implement in the above equations are
\begin{equation}
    \begin{split}
        u_{i,j}(\gamma) &= ZZ^{(J)}_{i,j}(\gamma+
        \frac{\pi}{4J_{i,j}})\;XY_{i,j}(\frac{\pi}{4}) , \\
        \tilde{u}_{i,j}(\gamma,\beta) &= ZZ^{(J)}_{i,j}(\gamma+\frac{\pi}{4J_{i,j}})\;XY_{i,j}(\beta+\frac{\pi}{4})
    \end{split} \ ,
\end{equation}
and  $U_{B}\rbracket{\beta}$ is a standard QAOA mixer consisting of $RX$ rotations.
Recall from Eq.~\eqref{eq:identity_swaps} that phase shifts in $ZZ$ and $XY$ gates allow to efficiently implement SWAPs without additional physical gates. 

We are thus interested in implementing gates of the form (omitting the dependence on $J_{i,j}$ for clarity)
\begin{equation}
ZZ_{i,j}\rbracket{\gamma} \coloneqq \exp\rbracket{-i \gamma\ Z_iZ_j} \ ,
\end{equation}
\begin{equation}
    XY_{i,j}\rbracket{\beta} \coloneqq \exp\rbracket{-i \beta\ (X_iX_j+Y_iY_j)} \ .
\end{equation}
In Rigetti's Aspen-M-3 device, XY gate is native, but the ZZ rotation needs to be decomposed further.
We use the decomposition
\begin{equation}
    ZZ_{i,j}\rbracket{\gamma} = \rbracket{RZ_i\rbracket{2\gamma} \otimes RZ_j\rbracket{2\gamma}}\ \text{CPHASE}_{i,j}\rbracket{4\gamma} \ ,
\end{equation}
where 
\begin{equation}
    RZ_i\rbracket{\gamma} = \exp\rbracket{-i\frac{\gamma}{2} Z_i} \ ,
\end{equation}
\begin{equation}
    \text{CPHASE}_{i,j}\rbracket{\gamma} = \text{diag}\rbracket{1,1,1,e^{-i\gamma}} \ .
\end{equation}
The CPHASE and RZ gates are implemented natively in Rigetti's hardware. 
Importantly, RZ gates are implemented virtually via signal frame changes in the classical controls \cite{mckay2017efficient}, which makes their implementation effectively error-free.

To summarize, the implementation of the $p$-layer time-block $k$-QAOA/QAMPA ansatz described in the text requires $\mathcal{O}\rbracket{pkn}$ two-qubit quantum gates and $\mathcal{O}\rbracket{pkn}$ single-qubit near-error-free RZ rotations (the exact number varies slightly depending on the parity of $\noq$ and $k$).
For QAMPA there are no additional gates, while for QAOA each layer is accompanied by single-qubit rotations around $X$ axis
\begin{align}
RX_i\rbracket{\beta}=\exp\rbracket{-i\frac{\beta}{2} X_i} \ ,
\end{align}
that are implemented using fixed calibrated $RX\rbracket{\frac{\pi}{2}}$ pulses combined with arbitrary $RZ$ rotations via identity
\begin{align}
    RX\rbracket{\beta} \approx RZ\rbracket{\frac{\pi}{2}}RX\rbracket{\frac{\pi}{2}}RZ\rbracket{\beta+\pi}RX\rbracket{\frac{\pi}{2}}RZ\rbracket{\frac{\pi}{2}} \ ,
\end{align}
with $\approx$ denoting equivalence up to a global phase.

\subsection{Choosing sublattices}
The Rigetti's Aspen-M-3 chip has 79 functional qubits. 
In our experiments, we used subsystems of smaller sizes, up to $\noq=50$. 
Recall that our ans\"atze require linear connectivity. 
In the case of Aspen-M-3, there are usually multiple choices for linearly-connected chains of qubits of size $\noq\leq 50$. 
To account for temporal changes in gate fidelities, at each calibration window (i.e., a period of time between two device's calibrations), we performed heuristic optimization over linear chains to find the sublattice of the highest effective fidelity. 
Specifically, we optimized a cost function that was a product of fidelities of all one- and two-qubit gates possible to implement in a sublattice. 
Since all of the experiments presented in this work were implemented across multiple days, this implies that different data points generally correspond to different physical sublattices.

\subsection{Hamiltonian ground state energies fluctuations}\label{app:sec:hamiltonians_gs}

\begin{table}[t!]
\begin{tabular}{|c|llll|}
\hline & \multicolumn{4}{c|}{\textbf{Ground state energy}} \\ \cline{2-5} 
\multirow{-2}{*}{\textbf{System size}} & \multicolumn{1}{c|}{min}                         & \multicolumn{1}{c|}{mean}                          & \multicolumn{1}{c|}{max}                         & \multicolumn{1}{c|}{CV}       \\ \hline
\textbf{30}                           & \multicolumn{1}{l|}{{\color[HTML]{242424} -129}} & \multicolumn{1}{l|}{{\color[HTML]{242424} -113.4}} & \multicolumn{1}{l|}{{\color[HTML]{242424} -103}} & {\color[HTML]{242424} -6.6\%} \\ \hline
\textbf{40}                           & \multicolumn{1}{l|}{{\color[HTML]{242424} -202}} & \multicolumn{1}{l|}{{\color[HTML]{242424} -177}}   & \multicolumn{1}{l|}{{\color[HTML]{242424} -166}} & {\color[HTML]{242424} -6\%}   \\ \hline
\textbf{50}                           & \multicolumn{1}{l|}{{\color[HTML]{242424} -261}} & \multicolumn{1}{l|}{{\color[HTML]{242424} -251.2}} & \multicolumn{1}{l|}{{\color[HTML]{242424} -243}} & {\color[HTML]{242424} -2.4\%} \\ \hline
\end{tabular}
\caption{Statistical properties of the Sherrington-Kirkpatrick Hamiltonian instances used in the experiment.
For each system size, the minimal, mean, maximal, and coefficient of variation (CV) of ground state energies are calculated over $10$ random instances used in our experiments.}\label{tab:app:hamiltonians_gs}
\end{table}

All experiments presented in the manuscript were performed on 10 randomly chosen instances of the Sherrington-Kirkpatrick model. 
Since the amount of tested instances is rather small, it is important to consider whether small-size effects might significantly affect our results. 
To this aim, in Table~\ref{tab:app:hamiltonians_gs} we gather some statistical information on ground state (GS) energies of the implemented instances.
We observe that fluctuations in our ensemble are fairly small, reaching coefficient of variation (CV) of $\approx 6\%$ for $\noq = 30$ and $\noq = 40$, and CV above $2 \%$ for $\noq = 50$.
Based on the above, we do not expect the GS energy fluctuations of the implemented instances to highly affect our results.
However, we note that in Section~\ref{app:sec:encodings} we identify another property (energy of $\ket{0\dots 0}$) of the Hamiltonians that likely did affect the performance due to higher spread between the instances.

\section{Additional details on results}\label{app:additional_results}

\subsection{Output Distribution Tails analysis}\label{app:results_tails}

\begin{figure*}[!h]
\begin{center}
{\includegraphics[width=1.0\textwidth]{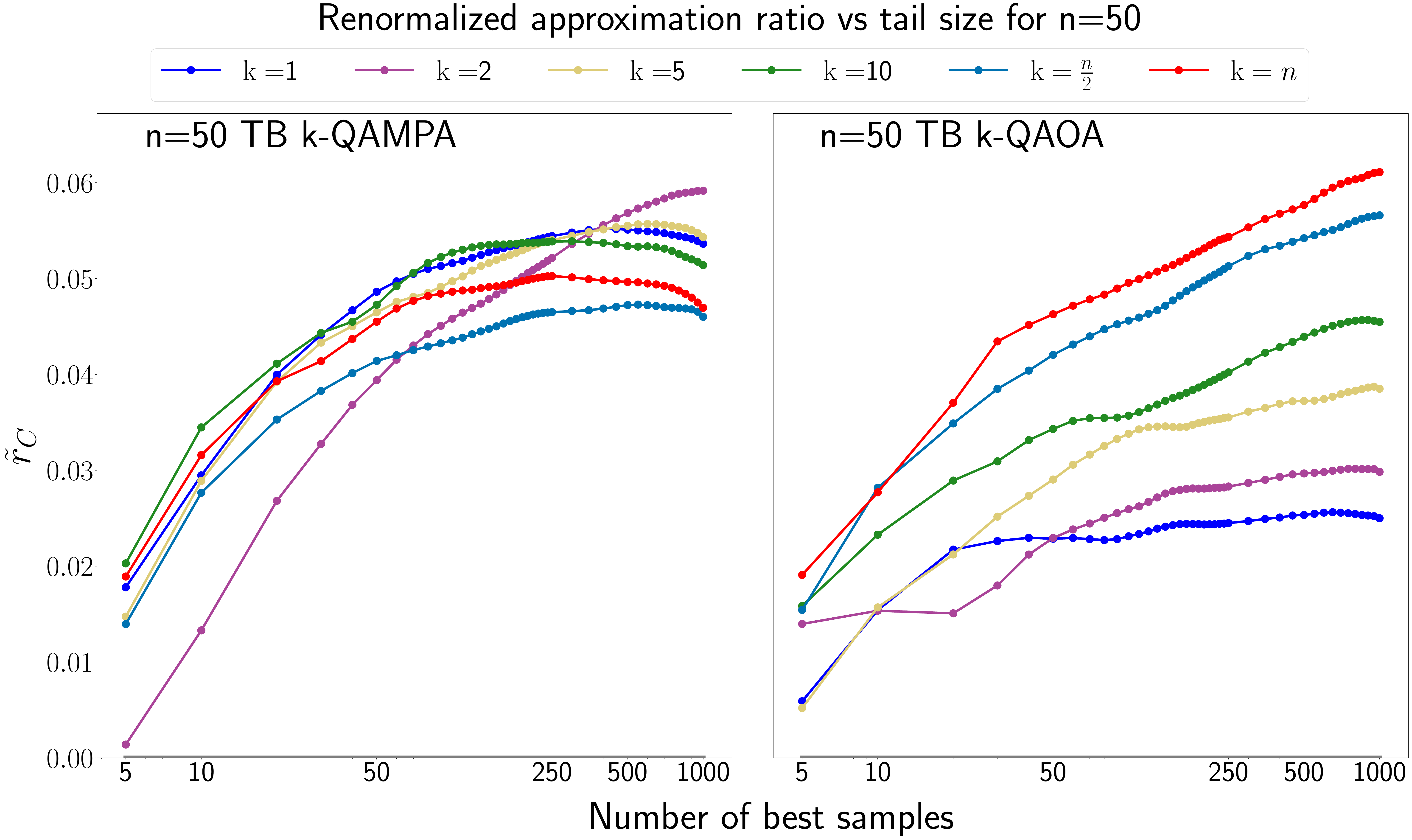}} 
  \caption{
  Average (over $\noH$ random Hamiltonian instances) expected value of the renormalized approximation ratio (Eq.~\eqref{eq:ar_renormalized}) calculated from a varying number of the lowest energy samples (Eq.~\eqref{eq:ar_quantile}).
  Note that due to renormalization, positive values correspond to better-than-random performance.
 For each $k$, each datapoint is mean over depths corresponding to range $\frac{pk}{n} \in \sbracket{1,2}$, i.e., between depth 1 and 2 of a standard ansatz.
  The rest of the plot's convention is the same as for Fig.~\ref{fig:best_results}.
  }
  \label{fig:tails_plot}
 \end{center}
\end{figure*}

Recall that to guide variational optimization, typically the standard estimator of approximation ratio is used (Eq.~\eqref{eq:ar_singleshot}), and that is what we presented in plots in the main text.
However, that estimator is not the only relevant figure of merit for assessing the quality of the optimization.
Indeed, one is generally interested in the single, best solution obtained.
Having optimized the cost function, one prepares a final variational state on a quantum device. 
 This allows for obtaining multiple samples from an optimized probability distribution and thus obtaining multiple candidate solutions.
A desirable property of the optimized probability distribution is to have support in the region with low-energy states.
If that is the case, sufficient sampling might allow to access the tails of the distribution that could, in principle, be close to the optimal solution.
 Hence it is 
 desirable to assess additional properties of the optimized distribution.
 
In view of future explorations quantifying the real-world performance of the optimization solver~\cite{windowsticker, lubinski2023optimization}, we now consider the tails of the estimated distributions by constructing estimators of the form 
\begin{equation}\label{eq:ar_quantile}
        \langle\tilde{C} \rangle
        = \frac{1}{\tilde{s}}\sum_{j=1}^{\tilde{s}}C_j \ ,
\end{equation}
where $\tilde{s} \in \cbracket{1,\dots,s}$, and the $\cbracket{C_j}_{j=1}^{s}$ have been sorted in 
ascending order.
Observe that choosing $\tilde{s}=s$ corresponds to the standard estimator, 
while $\tilde{s}=1$ corresponds to 
the minimum energy solution found among the outputs of the series of shots.

To study the above, we investigate how the relative advantage over random bitstrings sampling changes when one looks at multiple tails, i.e., by calculating approximation ratio average over $\tilde{s}$ (i.e., Eq.~\eqref{eq:ar_quantile} divided by ground state energy).
As we are interested in \emph{relative} improvement over the random baseline, we will consider the following renormalized approximation ratio
\begin{align}\label{eq:ar_renormalized}
    \tilde{r}_{\tilde{C}} = \frac{r_{\tilde{C}}^{\mathrm{exp}}-r_{\tilde{C}}^{\mathrm{random}}}{1-r_{\tilde{C}}^{\mathrm{random}}} \ ,
\end{align}
where $r_{\tilde{C}}^{\mathrm{exp}}$ denotes the value obtained experimentally, and $r_{\tilde{C}}^{\mathrm{random}}$ a value obtained via random sampling.
The above figure of merit is positive when the approximation ratio outperforms the white noise (random sampling), and 1 when the optimal solution is found (which corresponds to $r_{\tilde{C}}^{\mathrm{exp}}=1$).
To make the comparison with experiments fair, such random baseline estimators will be calculated using the same number of samples (from a uniform distribution) as the number of samples in our experiments.
For random baseline samples the sampling process is repeated multiple times and an average is taken to account for statistical fluctuations.

In Fig~\ref{fig:tails_plot}, for fixed $k$ and $\noq=50$, we look at the renormalized approximation ratio (Eq.~\eqref{eq:ar_renormalized}) averaged over the instances as well as over the (varying) amount of $\tilde{s} \in \cbracket{1,\dots,s}$ \emph{best} samples (recall Eq.~\eqref{eq:ar_quantile}).
Note that data points are easier to distinguish in a logarithmic scale.
Recall that for QAMPA ansatz for high system size, we observed a roughly monotonic increase in performance with circuit depth up until around $\frac{pk}{n}\approx 1$, after which the dependence seems to flatten (for QAOA the dependence was flat for all depths).
To account for fluctuations in the best points (corresponding to high depths), each data point in Fig.~\ref{fig:tails_plot} is an \emph{average} of data points corresponding to depths in range $\frac{pk}{n}\in \sbracket{1,2}$.

Interestingly, for QAMPA, for all $k$ we observe performance better than random for almost the whole range of tails from $\tilde{s}=5$ (i.e., top 2.5\% percentile) to $\tilde{s}=1000$ (i.e., standard approximation ratio). 
Furthermore, looking at the tails ($\tilde{s}\leq 50$) indicates that the results for $k=5$ for QAMPA ansatz either outperform or are on par with the results for $k=n$ (corresponding to standard ansatz) until around $\tilde{s}\approx 50$, where standard ansatz becomes the most performant.
This suggests that the distributions obtained from TB ansatz might have overall better properties than those obtained using standard ansatz. 
In the case of QAOA, we observe that all tails perform better than random.
We further observe that increasing $k$ increases the quality of all of the tails, indicating that for QAOA the TB ansatz performs worse than the standard ansatz. 

We note that the possibility of exploring a more fine-grained range of physical circuit depths is a significant advantage of the proposed ans\"atze over standard approaches -- especially when lower physical depth allows us to obtain the same or better performance, which we observed for both QAMPA and QAOA an\"satze for some data points.

\begin{figure*}[!h]
\begin{center}
\includegraphics[width=1.0\textwidth]{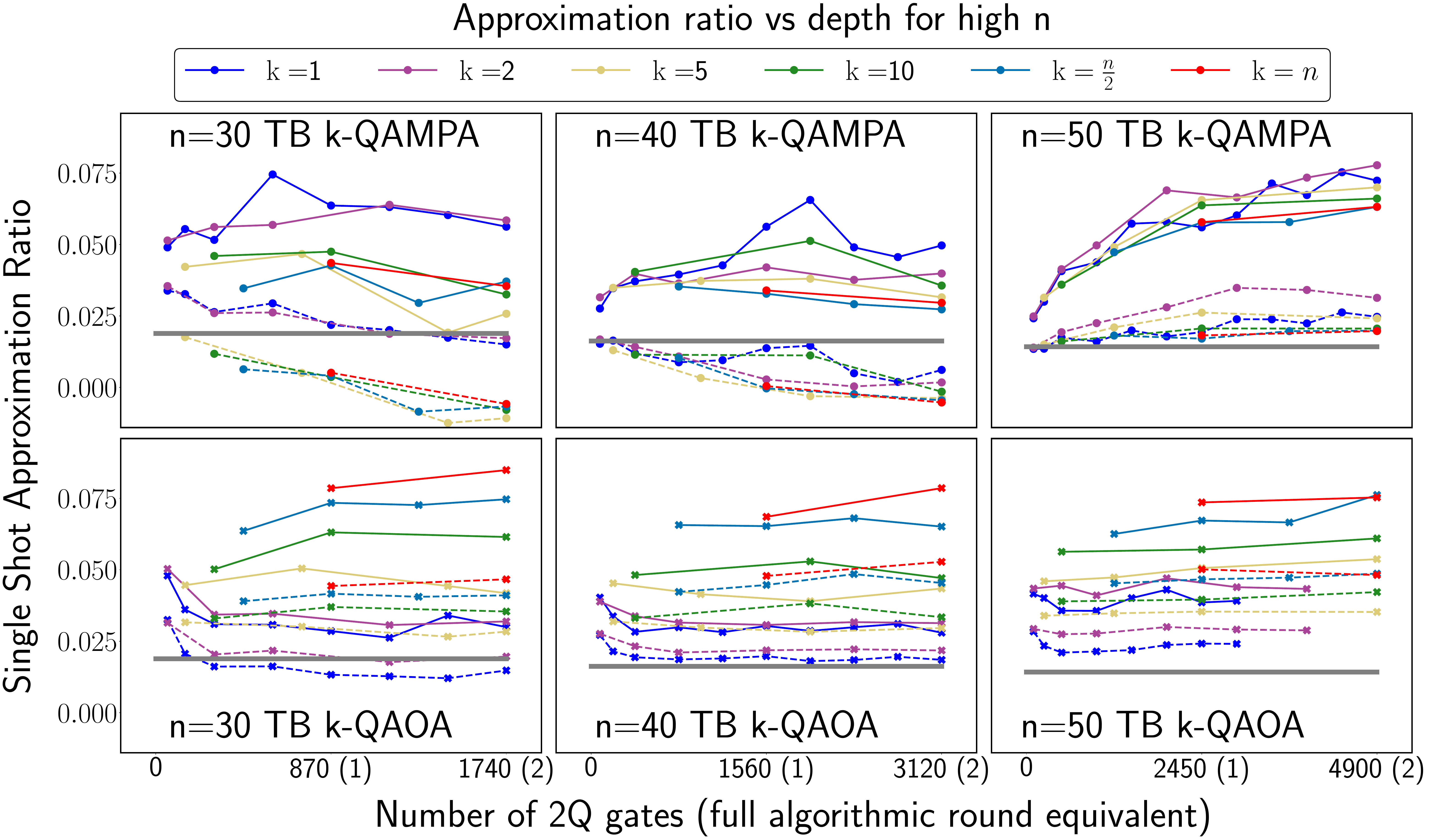} 
  \caption{
  Average (over $\noH$ random Hamiltonian instances) estimated expected value of the approximation as a function of circuit depth for QAMPA (top) and QAOA (bottom).
  Solid lines correspond to the best ansatz gate ordering, while dashed lines correspond to the average over 10 random gate orderings.
  The rest of the plot's convention is the same as for Fig.~\ref{fig:best_results} (which is zooming in on the rightmost column showing also adaptive parameter optimization results).
  }\label{fig:avg_vs_best_1000}
 \end{center}
\end{figure*}

\begin{figure*}[!t]
\begin{center}
\includegraphics[width=1.0\textwidth]{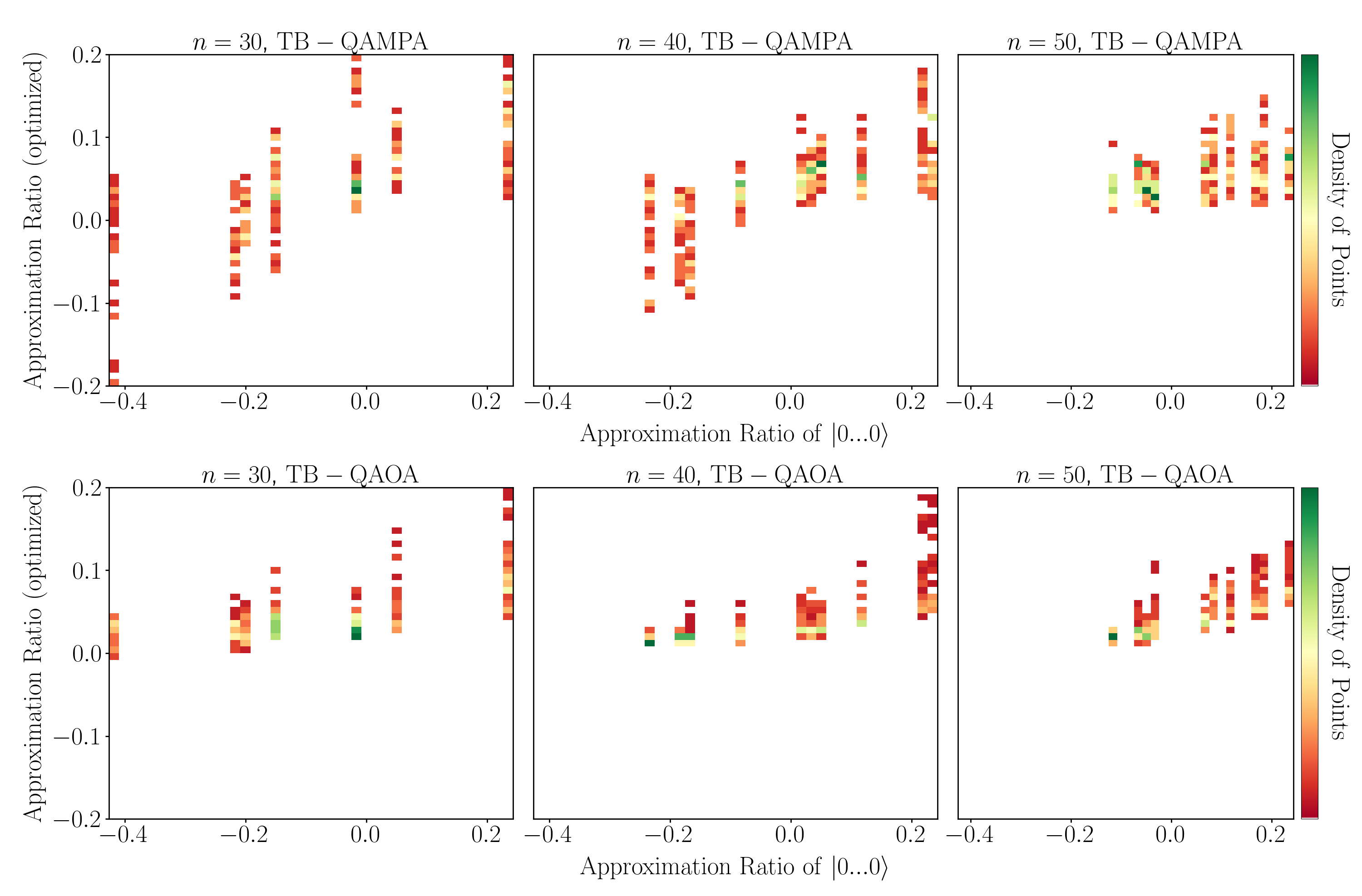}
  \caption{
  Correlation between approximation ratio of $\ket{0\dots 0}$ state (X-axis) and the approximation ratio obtained in optimization (Y-axis) using random optimizer (and the best gate ordering). 
  The first row of plots corresponds to Time-Block QAMPA and the second row to Time-Block QAOA. Columns correspond to different system sizes.
  For each subplot, the shown data is aggregated over all values of $k$ and $p$ and all Hamiltonian instances.
  For $n=50$, the shown results correspond to the datapoints from Fig.~\ref{fig:best_results} for best gate orderings.
  }
\label{fig:ar0_vs_ar}
 \end{center}
\end{figure*}

\begin{figure*}[!t]
\begin{center}
\includegraphics[width=1.0\textwidth]{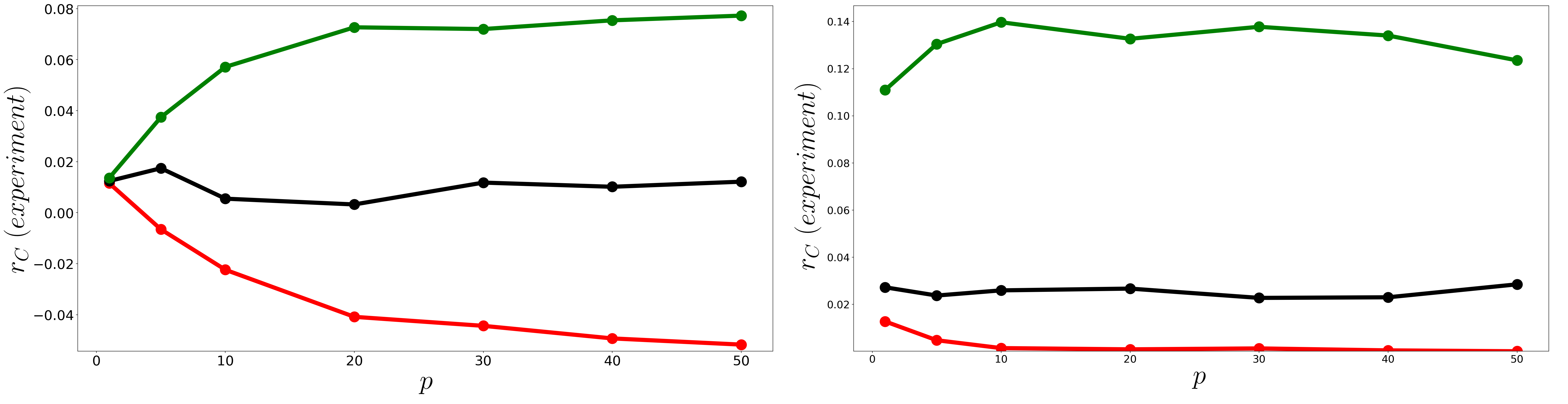}
  \caption{
  Investigation of implementing bitflip transform pre-processing on the performance of the Time-Block $k=2$ QAMPA (left) and QAOA (right) ansatze.
  The X-axis shows depth, while the Y-axis corresponds to the approximation ratio obtained in experiments with random angles optimizer. 
  The green lines correspond to bitflip transforms that are expected to perform well, black to medium performance, and red to transforms that are expected to perform badly.
  See text description for details.}\label{fig:preprocessing_gauges}
 \end{center}
\end{figure*}

\subsection{Performance for smaller system sizes}\label{app:sec:results_small_n}

\subsubsection{Performance trends}
In Fig.~\ref{fig:avg_vs_best_1000} we present performance overview plots analogous to Fig.~\ref{fig:best_results} in the main text but extended by system sizes of $\noq=30$ and $\noq = 40$.
The experiments were performed over range of $k$s and $p$s using random angles optimizer for the same class of random SK Hamiltonians, and on the same device, Aspen-M-3.
It is notable that the performance does not seem to drop with the system size.
Moreover, for best gate orderings, the trend of increase of performance with depth holds for QAMPA only for $\noq=50$, with $\noq=40$ being roughly flat, and $\noq=30$ exhibiting a drop of performance with depth.
The drop of performance with depth is particularly striking for average gate orderings for $\noq =30, 40$. 
This is an unexpected behavior, as one expects the optimization to perform worse with growing system size.
We attempt to formulate a partial explanation of this effect in the next subsection.

\subsubsection{Effects of mapping on optimization performance}\label{app:sec:encodings}

To partially explain the reverse trend from Fig.~\ref{fig:avg_vs_best_1000}, we follow Ref.~\cite{maciejewski2024ndar}, where the authors demonstrated a strong dependence between a Hamiltonian mapping and QAOA performance. 
Here by mapping we mean the formal association of $\ket{0}$ and $\ket{1}$ states with $\pm 1$ eigenvalues of Pauli $Z$ operator.
Note that such association can be changed on the level of the Hamiltonian operator by applying a change-of-basis unitary transformation (a \emph{bitflip transform}) corresponding to bitflip on qubits for which we wish to swap the association.
Some types of noise, such as amplitude damping, exhibit strong bias towards one of the computational basis states on each qubit.
This causes some states to be favored by noise.
In particular, for the amplitude damping channel, the global ``attractor’’ state is $\ket{0\dots 0}$.
As noted in Ref.~\cite{maciejewski2024ndar}, one might thus expect that optimization with Hamiltonian mapping for which $\ket{0\dots 0}$ state has high approximation ratio will perform better.
This is because, in a strong noise regime, the effective space attainable by a given ansatz will be, in general, reduced to some proximity of the attractor state.
While experiments in Ref.~\cite{maciejewski2024ndar} were implemented on the newest generation of Rigetti's chip Ankaa-2, we expect similar noise characteristics for older Aspen-M-3 used in our experiments. 

To investigate this effect, we first follow Ref.~\cite{maciejewski2024ndar} by performing the following data analysis.
Note that every data point from Fig.~\ref{fig:avg_vs_best_1000} is characterized by a tuple $\left(\textit{base\ ansatz}, n, k, p, i\right)$, where $\mathrm{base\ ansatz}$ is the original ansatz from which Time-Block ansatz was constructed (QAOA or QAMPA), $n$ denotes system size, $k$ corresponds to $k$-Time-Block ansatz, $p$ is algorithmic depth, and $i$ is Hamiltonian index.
For each system size, we now consider every Hamiltonian instance and calculate 
\begin{align}\label{app:eq:ar_zero}
    r_{0} = \frac{\bra{0\dots 0}H\ket{0 \dots 0}}{C_{\min}} \ ,
\end{align}
i.e., the approximation ratio of postulated noise attractor state $\ket{0\dots 0}$.

Now for each $\textit{base\ ansatz}$ and system size $n$, we take all data points (for every $k$, $p$, and $i$), and plot the approximation ratio attained via optimization vs approximation ratio of zero state from Eq.~\eqref{app:eq:ar_zero}. 
The 2-dimensional histograms are shown in Fig.~\ref{fig:ar0_vs_ar}.
A fixed value of $r_0$ (X-axis) in the plot corresponds to Hamiltonians (or a single Hamiltonian) with a given AR of $\ket{0\ \dots 0}$.
We make the following observations.
First, for $\noq = 30$, the spread of $r_{0}$ values is the highest. 
In particular, there are Hamiltonian instances for which $r_0$ is much smaller than $0.0$ (up to $\approx -0.4)$.
For $\noq = 40$ we observe a smaller spread, but many instances also exhibit values much smaller than $0.0$ (up to $\approx -0.2$).
On the other hand, $\noq=50$ exhibits the smallest spread, with majority of $r_0$ values being higher than $0$, and those with negative values having relatively small magnitude (up to $\approx -0.1$).
Second, we observe that for QAMPA and QAOA with system size $\noq = 30$, and less so with $\noq = 40$, there exists a rough positive correlation between $r_0$ and the optimization performance.
Notably, for QAMPA the correlation is visibly stronger.
For $\noq = 50$, the correlation is less clear for both ansatze (as mentioned above, the spread over $r_0$s for that system size was the smallest, which can partially explain this).
However, for all system sizes (including $\noq = 50$), we note that the best optimization performance corresponds to higher values of $r_0$, and the worst performance to lower values of $r_0$.
Finally, we note that for QAMPA, the spread in overall approximation ratio values ($Y$-axis on Fig.~\ref{fig:ar0_vs_ar}) is much higher for $\noq = 30$ and $\noq=40$ than for $\noq = 50$, which correlates with the above-described trends.
In particular, there are multiple particularly bad-performing instances for QAMPA with $\noq=30$ and $\noq=40$, with low values of $r_0$. 

The above analysis suggests a possible explanation of the reverse depth vs performance trends (w.r.t. system size) for QAMPA observed in Fig.~\ref{fig:avg_vs_best_1000}.
Namely, the default mappings of the Hamiltonian instances that we benchmarked turned out to be particularly adversarial w.r.t. noise characteristics for smaller system sizes. 
To support this explanation further, in the next subsection, we perform experiments that aim to illustrate the magnified effects of mapping choice.

\subsubsection{Improving performance via mapping selection}

In the previous section, we demonstrated experimental evidence that the performance of the optimization (for a given instance) might correlate with the approximation ratio $r_0$ of $\ket{0\dots 0}$ (Eq.~\eqref{app:eq:ar_zero}) .
As proposed in Ref.~\cite{maciejewski2024ndar}, this fact can be exploited by changing the mapping of the Hamiltonian in a way that improves $r_0$.
The mapping change can be done in pre-processing using bitflip transforms, see Ref.~\cite{maciejewski2024ndar} for detailed discussion.
To investigate this, we implemented Time-Block $k=2$ QAMPA and QAOA ansatz for $\noq=50$ over different depth values with multiple bitflip transforms.
We first generated $10^6$ random transforms, and chose 5 transforms with the highest $r_0$ (expected to perform relatively well in optimization), 5 with the lowest $r_0$ (expected to perform relatively badly), and 1 random (expected to not affect performance much).

The results averaged over 10 random SK model instances are shown in Fig.~\ref{fig:preprocessing_gauges}.
In that figure, the green lines correspond to an average over results for $5$ best transforms (high $r_0$), black for $1$ random transform (random $r_0$), and red for $5$ worst transforms (low $r_0$).
Notably, we observe that for QAMPA, choosing favorable bitflip transforms allows to attain higher approximation ratios in optimization, as well as achieve the increase of the performance with depth trend.
Similarly, adversarial bitflip transforms reduce the performance and invert the trend.
On the other hand, QAOA seems to be less prone to the effects of mapping choice in terms of the depth vs performance trend, but absolute performance can be highly improved by the favorable bitflip transform choice.

\subsection{Parameter setting}\label{app:sec:results_angles_permutations}

\subsubsection{Relevance of gate orderings}

Recall that our ansatz has a freedom of gate ordering that leads to inequivalent logical circuits.
When implementing the random optimizer, we implemented each set of $100$ angles with $10$ random gate ordering (recall solid and dashed curves in Fig.~\ref{fig:best_results}).
To illustrate the importance of gate ordering optimization, in Fig~\ref{fig:avg_vs_best_1000} we plot the results for the best gate ordering next to the \emph{average} gate ordering, for $\noq\in\cbracket{30,40,50}$.
The results suggest that optimization of gate ordering is instrumental in obtaining a good performance for QAMPA ansatz.
For QAOA, it has less dramatic effects on the results, but it still can provide an advantage.
This difference is not surprising, as for QAOA, the logical circuits are more similar for different gate orderings. Indeed, for standard QAOA ansatz $k=\noq$ all gate orderings are nominally identical because the two-qubit gates commute. However, results can still differ in practice due to noise effects.

\subsubsection{Relative importance of gate orderings vs angle optimization}

To compare the relative importance of the choice of angles and gate orderings, we perform the following analysis.
For each triple $(\noq,k,p)$, we look at the approximation ratio for each set of angles, gate ordering, and Hamiltonian.
For each Hamiltonian, we first loop over gate orderings, and calculate 
a maximal spread of the optimized function (in this case the approximation ratio estimator) over all range of implemented angles. 
We then average over gate orderings and over the $\noH$ random Hamiltonians.
Explicitly, treating $r_C$ as a function of gate orderings and angles, for fixed Hamiltonian we calculate
\begin{align}
\delta_{\mathrm{max}}\rbracket{i} & \coloneqq \max_{\alpha,\beta} r_C(i,\alpha)-r_C(i,\beta) \ ,\\
    \delta_{\mathrm{max}} &= \left<\delta_{\mathrm{max}}\rbracket{i}\right>_{i} \ ,
\end{align}
where index $i$ labels a gate ordering and $\alpha,\beta$ are variational angles, and we average the above quantity $\delta_{\max}$ over Hamiltonian instances.

\begin{figure*}[!h]
\begin{center}
{\includegraphics[width=\textwidth]{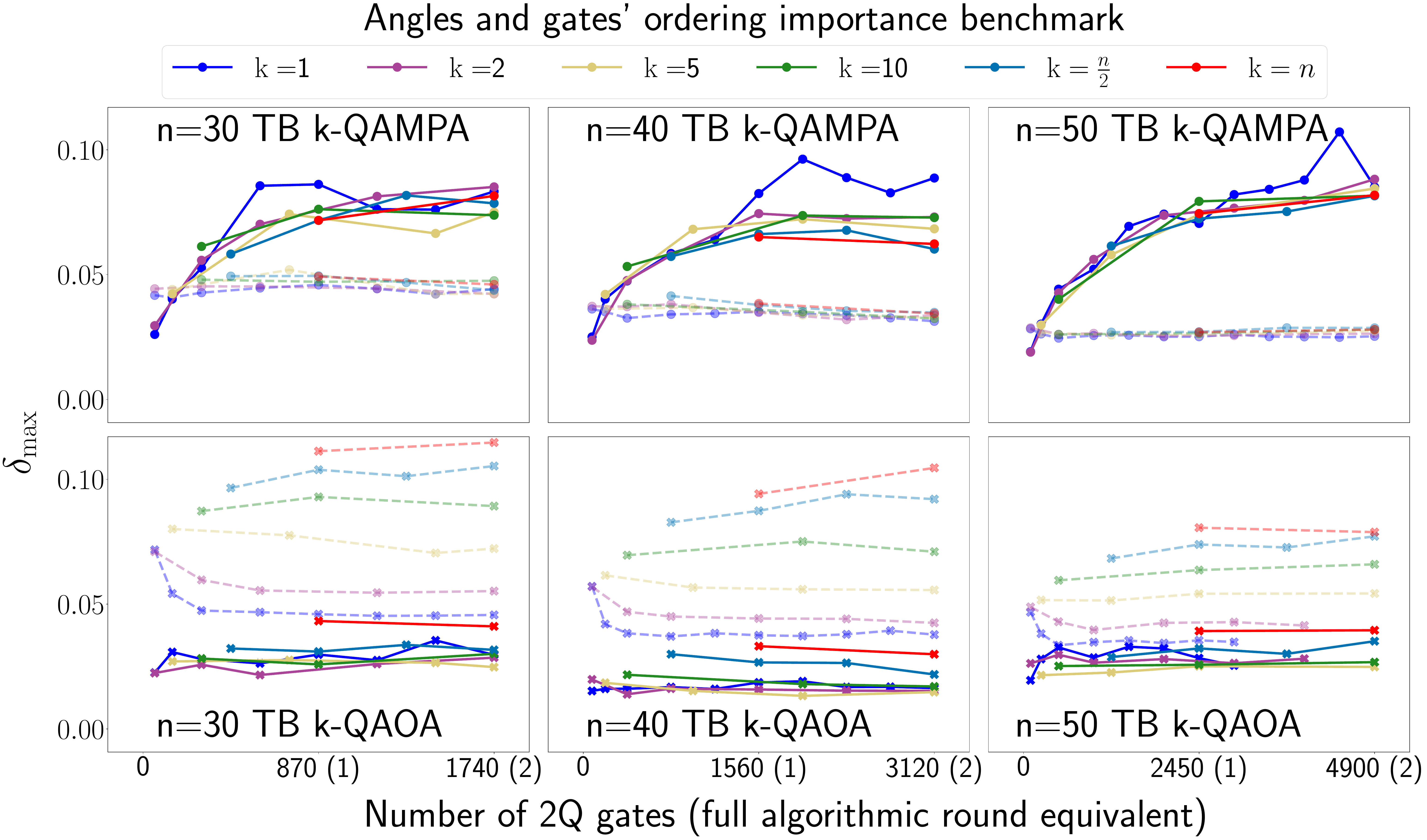}}
  \caption{
  Mean (over second variable) of maximal spreads (over first variable) of the optimized function value, when varying the first variable for the fixed second variable.
  Here variables correspond to gate orderings and variational angles -- solid lines represent data points where gate orderings are varied for fixed angles, dotted lines the reverse.
  }
  \label{fig:expected_spreads}
 \end{center}
\end{figure*}

Looking at the range of maximal spreads of the function value, allows us to roughly estimate how big an advantage can optimization of angles provide for fixed gate ordering.
We repeat the whole algorithm reversing the roles of gate orderings and angles, thus obtaining a maximal improvement when changing the gate ordering.
The results are plotted in 
Fig.~\ref{fig:expected_spreads}. 
For QAMPA, the plots suggest that optimization over gate orderings is important in the following sense.
For fixed angle, the average maximal improvement when varying gate ordering is much higher than the reverse (solid lines are always much above dotted lines in figures on the top).
Furthermore, we observe that the maximal spread when varying gate orderings grows roughly monotonically with circuit depth.
This is to be expected, as for higher circuit depth, the change of gate ordering changes the circuit more significantly.
For QAOA, on the other hand, the results suggest the opposite -- the gate ordering does not play as significant a role as for QAMPA, and angles' optimization is more important.

\subsubsection{Details of TPE optimization}\label{app:TPE}
To implement Tree of Parzen Estimators optimization \cite{Bergstra2011,hyperopt2013}, we used TPE implementation in the Python package \texttt{optuna} \cite{optuna2019} with version 3.1.0.
All hyperparameters were set to default values, with the exception that we turned off pruning trials, and we set \textit{n\_jobs}=4 to allow for parallel implementation over 4 threads.
Each optimization was run with a unique (random) seed for the optimizer, and each optimization started from the same initial points (all angles set to $0.1$).
Besides continuous angle variables, the optimizer was allowed to suggest categorical variables that represented one of the $50$ pre-generated random gate orderings.
Each optimization was run with $t\approx 800$ function calls.

\end{document}